**Amplified Summer Wind Stilling and Land Warming Compound Energy Risks in Northern Midlatitudes**

Gan Zhang[1*]

[1]Department of Climate, Meteorology, and Atmospheric Sciences,

University of Illinois at Urbana-Champaign, 1301 W. Green Street, Urbana, IL 61801

*Corresponding Author: Gan Zhang

Email: gzhang13@illinois.edu

**Abstract**

Wind energy plays a critical role in mitigating climate change and meeting growing energy demands. However, the long-term impacts of anthropogenic warming on wind resources, particularly their seasonal variations and potential compounding risks, remain understudied. Here we analyze large-ensemble climate simulations in high-emission scenarios to assess the projected changes in near-surface wind speed and their broader implications. Our analyses show robust wind changes including a decrease of wind speed (i.e., stilling) up to ~15% during the summer months in Northern Midlatitudes. This stilling is linked to amplified warming of the midlatitude land and the overlying troposphere. Despite regional and model uncertainties, robust signals of warming-induced wind stilling will likely emerge from natural climate variations in the late 21$^{st}$ century of the high-emission scenarios. Importantly, the summertime wind stilling coincides with a projected surge in cooling demand, and their compounding may disrupt the energy supply-demand balance earlier. These findings highlight the importance of considering the seasonal responses of wind resources and the associated climate-energy risks in a warming climate. By integrating these insights into future energy planning decisions, we can better adapt to a changing climate and ensure a reliable and resilient energy future.

# 1. Introduction

In the face of climate change, wind energy has emerged as a crucial solution for decarbonizing electricity generation. Over the past decades, wind energy grew rapidly thanks to decreasing project costs and favorable policy and investment environments (Bilicic and Scroggins 2023). As the world strives for net-zero emissions, the continued rise of greenhouse gases due to human activities may pose new challenges to the electricity system. Besides increasing the cooling demand, global warming also drive profound changes in atmospheric circulation patterns. These changes include warming-induced weakening of circulation strength (Held and Soden 2006, Coumou *et al* 2015), latitudinal shifts of midlatitude westerlies (Lorenz and DeWeaver 2007, Simpson *et al* 2014), and seasonality shifts (Huang *et al* 2013, Song *et al* 2018, Zhang 2023), all of which can affect the spatial-temporal distribution of wind resources. These changes, in turn, pose potential challenges to electricity grids that are increasingly reliant on wind energy. These challenges may also complicate efforts to meet demand changes driven by non-climate factors (e.g., sector electrification and data centers) (International Energy Agency 2024). Therefore, understanding and proactively addressing the climate-driven risks associated with wind energy is crucial for ensuring a smooth and cost-effective energy transition.

The past and future changes in the near-surface wind speed have received intense interest, but their causes and robustness are still debated (Pryor *et al* 2020, Yalew *et al* 2020, Jung and Schindler 2022) due to conflicting observations and model projections. Observations suggest that the near-surface wind speed reduced significantly between 1980 and 2010 in the northern hemisphere (NH) midlatitudes (Vautard *et al* 2010, Torralba *et al* 2017). Hydrology researchers noticed this decrease in wind speed and termed it as terrestrial wind stilling (McVicar *et al* 2012 and references therein). This stilling term was later adopted by hydrology and wind energy studies (e.g., Vautard *et al* 2010,

Zeng *et al* 2019, Deng *et al* 2022). Various explanations of wind stilling have been proposed (McVicar *et al* 2012), including vegetation changes (Vautard *et al* 2010) and circulation responses to aerosol and greenhouse-gas forcings (e.g., Karnauskas *et al* 2018, Shen *et al* 2021). However, the long-term stilling trend experienced a reversal in the 2010s (Zeng *et al* 2019), potentially due to natural climate variability (Wohland *et al* 2021, Deng *et al* 2022) and changes in record-keeping (Dunn *et al* 2022). Regarding future wind resources, existing regional studies suggest small or mixed changes (Solaun and Cerdá 2019, Pryor *et al* 2020, Yalew *et al* 2020), but recent studies of global climate model (GCM) simulations show that anthropogenic warming can result in significant, continent-scale wind changes (Karnauskas *et al* 2018, Zha *et al* 2021, Deng *et al* 2022). Nonetheless, the direction of future changes remains a significant point of contention, even for extensively studied regions like the UK (Hosking *et al* 2018, Karnauskas *et al* 2018, Deng *et al* 2022).

Those disagreements appear to partly arise from unforced, natural climate variability. Except for a few recent studies (Wohland *et al* 2021, Deng *et al* 2022, Zha *et al* 2021), nearly all the existing studies of future wind speed drew their conclusions by analyzing a single run or small ensembles (≤ 5) of regional climate simulations (Pryor *et al* 2020). These regional studies often rely on downscaling techniques, which use coarse-resolution GCM simulations as input and simulate regional changes at higher resolutions. While the high resolutions help model terrain impacts on wind, the small ensemble size make it difficult to determine to what extent the simulated changes were driven by unforced natural variability or anthropogenic warming (Wohland *et al* 2021). Furthermore, some regional studies exclusively rely on regional climate simulations and show a distinct disconnection from the extensively studied changes in large-scale atmospheric circulation (Jung and Schindler 2022). These issues revolving around simulations and physical interpretations ultimately limit confidence in the projected changes.

While the projected wind changes appear small (<5%) on an annual average basis (Pryor *et al* 2020, Yalew *et al* 2020, Jung and Schindler 2022, Wohland *et al* 2021, Deng *et al* 2022, Zha *et al* 2021), research on large-scale atmospheric circulation (Coumou *et al* 2015, Simpson *et al* 2014, Deng *et al* 2022) suggests that warming-induced wind changes are seasonally amplified. Few studies on wind energy have considered this seasonal amplification and its potential to exacerbate risks for electricity grids. Since wind turbine power output approximately scales with the cube of wind speed, even a moderate reduction (e.g., 5-10%) in wind speed can substantially reduce the electricity output of wind turbines. Such a reduction may have important implications for wind energy investments and the resource adequacy for electricity grids, potentially leading to power shortages in critical circumstances (e.g., heatwaves). A combination of low-wind and heatwave conditions may manifest as a compound risk for future societies (Zscheischler *et al* 2018).

Although significant gaps remain between current GCM simulations and the requirements of energy applications (Craig *et al* 2022), GCM simulations will likely remain essential for understanding large-scale atmospheric changes and continue serving as input for the regional downscaling of wind resources. We posit that a deeper understanding of GCM projections and their implications for wind energy is valuable. In this study, we analyze large-ensemble GCM simulations to identify robust, warming-induced changes in wind resources. We highlight the pronounced wind stilling in the summertime of the NH midlatitudes. Our seasonal perspective also reveals inconsistencies between the projected wind stilling and previous physical interpretations and suggests the importance of amplified warming of the midlatitude land. Lastly, we examine the compounding risk related to declining wind power output and surging cooling demand. These results have important societal implications in the high-emission scenarios and suggest mitigating the future climate-energy risk requires diversified, complementary energy sources.

## 2. METHODS

### 2.1 Historical Data and Climate Simulations

Our historical climate reference is the fifth-generation atmospheric reanalysis by the European Centre for Medium-Range Weather Forecasts (ECMWF), commonly known as the ERA5 (Hersbach *et al* 2020). The ERA5 is an observation-constrained model estimate of atmospheric states. The spatial resolution of the raw ERA5 data is ~0.25 degrees. While the ERA5 has limitations at fine spatial-temporal scales, it nonetheless provides useful information and outperform similar datasets (Olauson 2018, Molina *et al* 2021). The analyses of the ERA and climate simulations use the monthly 10-meter wind speed data, which were calculated with intraday, high-frequency wind speed by data providers (e.g., ECMWF). The choice of analyzing monthly mean data is due to the lack of high-frequency, turbine-level wind data output in many climate simulations. The limitations and basis of using monthly wind data to estimate wind energy potential are summarized in Supplementary Materials. While the lack of turbine-level data is also challenging, a set of assumptions can help establish that the relative changes in wind speed are equal at 10-meter and arbitrary turbine levels (Supplementary Materials). To highlight key regions for wind energy, we also use the wind farm data from the Global Energy Monitor (2023), which provide information of global wind energy projects.

Our primary climate simulation dataset is the large ensemble simulation by the Community Earth System Model 2 (CESM2) (Danabasoglu *et al* 2020, Rodgers *et al* 2021). We use a 50-member subset with the smoothed biomass burning fluxes. The nominal resolution of this large ensemble simulation is 1-degree and finer than the resolution of most recent GCM-based wind energy studies (Karnauskas *et al* 2018, Zha *et al* 2021, Wohland *et al* 2021, Deng *et al* 2022). To evaluate the sensitivity of wind projections to model physics, resolution, and emission scenarios, we analyze two additional set of large-ensemble simulations, the Seamless System for Prediction and Earth System Research (SPEAR) (Delworth *et al* 2020) and the EC-Earth (Wyser *et al* 2021). The SPEAR and EC-Earth were developed independently of CESM2 and have higher spatial resolutions (~0.5 degrees). The details and performance of these models are available in Supplementary Materials and the model documentation papers.

Following the protocol of the Coupled Model Intercomparison Project phase 6 (CMIP6) (Eyring *et al* 2016), the experiments with CESM2, SPEAR, and EC-Earth simulate the historical climate up to 2014 and the future climate starting in 2015. The examined simulations used two warming scenarios characterized by the Shared Socioeconomic Pathways (SSPs) (O'Neill *et al* 2017). Specifically, the CESM2 ensemble uses SSP 3-7.0, and the SPEAR ensemble uses SSP 5-8.5. The EC-Earth experiments examined a range of SSPs, but we focus on the SSP 3-7.0 and the SSP 5-8.5 to facilitate comparisons with the CESM2 and SPEAR simulations. We acknowledge that the SSP 3-7.0 and the SSP 5-8.5 scenarios may not materialize owing to mitigation efforts (Supplementary Materials). Nonetheless, the strong forcings help discern warming-induced changes and estimate their upper bounds.

We also analyze CMIP6 simulations and prioritize twelve models that have relatively high spatial resolutions (see Supplementary Materials). Three models, GFDL-CM4, EC-Earth3, and NCAR-CESM2, are closely associated with the models used to generate the large-ensemble simulations described earlier. The CMIP6 and large-ensemble simulations used standard SSP forcings, except that the CESM2 large-ensemble simulations used smoothed biomass burning over 1990–2020. Considering the relatively small CMIP6 ensemble, we use the simulations with SSP5-8.5 forcings so that the simulated changes are larger and easier to detect. To facilitate the computation of multi-model means, we regridded all the CMIP6 simulation data to a 1-degree grid using the conservative method of the Earth System Modeling Framework (ESMF).

**2.2 Time of Emergence**

The time of emergence (ToE) is the time when a signal of climate change emerges from the noise of natural climate variability (Hawkins and Sutton 2012). Warming-induced changes may gradually strengthen and affect wind projects at different time horizons. We compare the 30-year climate of historical (1981–2010), near-future warming (2021–2050), and far-future warming (2071–2100) simulations. As warming-induced changes in the ensemble mean strengthen with time, they eventually exceed the natural variability range in the reference climate (1981–2010). To highlight the regions where distinguishable changes may appear sooner, we evaluate the time of emergence for regional averages and at grid points. Specifically, we follow past studies of estimating the time of emergence (Hawkins and Sutton 2012) and calculate the linear trend in wind speed of the ensemble mean between 2010 and 2100. We calculate the standard deviations ($\sigma$) in the reference simulations and use $\pm 2\sigma$ as the threshold of significant changes. Starting from 2010, the first time when the long-term changes meet this threshold is defined as the time of emergence. The same calculation is also conducted for individual ensemble members to estimate the ToE uncertainty (see Supplementary Materials).

**2.3 Degree Days**

The degree day metric is often used to estimate the energy demand for regulating indoor temperature (Miranda *et al* 2023, Staffell *et al* 2023). While unmet heating demand can lead to disastrous outcomes (Busby *et al* 2021), anthropogenic warming generally reduces the heating demand and increases the cooling demand. Accordingly, our analysis of climate-energy risk focuses on the cooling demand and investigate its changes using a simple definition of degree days:

$$CDD = \begin{cases} T_{avg} - T_{ref}, & T \geq T_{ref} \\ 0, & T < T_{ref} \end{cases}$$

Where $CDD$ is cooling degree days, $T_{avg}$ is the daily average of outdoor air temperature, $T_{ref}$ is a reference temperature. In practice, $T_{avg}$ is often defined as the average of maximum and minimum daily temperatures. $T_{ref}$ values depend on the user practice and often range between 18ºC and 26ºC. For simplicity, we use 18ºC for our primary analysis and present the sensitivity analysis of using different $T_{ref}$ values in Supplementary Materials. Here, the CDDs serve as a simple, conceptual proxy to compare energy demand across climates. Modeling the actual energy consumption needs to consider the impacts of additional factors (Staffell *et al* 2023).

## 3. RESULTS

### 3.1 Projections of Surface Temperature and Wind Speed

We first examine the distribution of wind energy development and the climate simulated by CESM2 (SSP3-7.0). Wind energy has been growing rapidly in the midlatitude regions of North America, Europe, and Asia (Figure 1a and Fig. S1), especially near the population centers (Fig. S2) where electricity demands for indoor temperature regulation and other uses are high. These midlatitude regions are characterized by pronounced seasonality of atmospheric circulation and surface temperature. The CESM2 large-ensemble simulations replicate the historical, large-scale climate patterns of near-surface wind (Fig. S3) and temperature (Rodgers *et al* 2021), despite biases such as overestimating ocean wind speeds in the springtime. This overall consistency supports the use of CESM2 large-ensemble simulations for investigating large-scale temperature-wind changes.

Comparing the late-century (2071–2100) and the historical (1981–2010) periods in the CESM2 simulations, the most pronounced surface warming occurs in the NH extratropics and shows strong seasonal variations (Figures 1b and 1c). While the wintertime warming is more pronounced in the NH polar regions, the summertime warming is more pronounced over the NH midlatitude land. The tropical surface warming also tends to be relatively strong over land. These warming patterns are salient in climate models and consistent with past studies on the polar (Manabe and Stouffer 1980, Holland and Bitz 2003, Stuecker *et al* 2018) and terrestrial (Sutton *et al* 2007, Byrne and O'Gorman 2018) amplification of anthropogenic warming. These patterns are crucial for understanding the projected wind changes and associated energy risks, as to be discussed later.

The CESM2 simulations also project significant changes in near-surface wind speed with distinct spatial and seasonal patterns (Figures 1b and 1c). The most pronounced regional changes occur in the tropical (e.g., Central Africa in Figure 1b) and the polar regions (e.g., the Arctic in Figure 1b), but most of these regions are unsuitable for large-scale wind energy development due to geographical or consumption constraints. Therefore, our subsequent analysis will emphasize the NH midlatitude regions with concentrated wind energy development (Figure 1a). In the NH wintertime, the projected wind changes over midlatitude land are weak except for a few regions (e.g., the Mediterranean region). In comparison, the summertime changes are widespread and substantial across the NH midlatitude land, with reductions generally around -10%. The overall robustness of these patterns is corroborated by the CMIP6 ensemble (Fig. S4).

The simulated changes in near-surface wind speed are closely related to the large-scale atmospheric circulation changes (Figures 2a–c). A robust atmospheric response to anthropogenic warming is the poleward shift of midlatitude westerlies (Lorenz and DeWeaver 2007, Simpson *et al* 2014). On an annual average, the core of the westerlies is located around 50ºS and 50ºN (Figure 1a). The westerlies shift poleward as the wind weaken on their equatorward side and strengthen on their poleward side (Figure 2a). The shifts in the surface westerlies are accompanied by similar shifts in the free troposphere (Figures 2b–c). Importantly, the seasonal decomposition reveals that the relatively weak wind changes in the NH annual mean are due to opposing seasonal changes that partially cancel each other out. Unlike the southern hemisphere (SH), the NH experiences a broad weakening of the midlatitude westerlies in the summertime, which is consistent with a weaker mean circulation and reduced synoptic-scale weather variability (Coumou *et al* 2015, Chang *et al* 2016). These seasonal circulation changes, along with the associated temperature changes, are summarized in Figure 2d.

Compared to the wintertime poleward shift of westerlies, the summertime wind stilling in the NH received less attention. Past studies have proposed several physical mechanisms to explain the summertime stilling. Specifically, these hypotheses include: i) the polar amplification hypothesis, which suggests that reduced meridional temperature (and pressure) gradients due to enhanced polar warming weakens the westerlies (Coumou *et al* 2015, Karnauskas *et al* 2018); ii) the surface roughness hypothesis, which attributes the wind stilling to increased surface roughness due to vegetation changes (Vautard *et al* 2010); and iii) the stability-mixing hypothesis, which posits that increased atmospheric stability hinders the downward mixing of momentum, leading to weaker surface winds (Deng *et al* 2022). However, a close inspection of the CESM2 simulations reveals inconsistencies between these mechanisms and the simulated wind changes (Supplementary Materials).

We propose a new mechanism to explain the summertime wind stilling in the NH, focusing on the role of amplified land warming. The NH has greater land coverage than the SH and warms more due to the terrestrial amplification (Figures 2b and 2c). Due to relatively active summertime convection, surface warming tends to extend to the troposphere, even in the midlatitudes (Zhang and Boos 2023). This leads to a vertically coherent warming pattern (Figure 2c), particularly between 40ºN and 75ºN where the land warming is pronounced (Figure 1c). Importantly, this tropospheric warming reduces the meridional temperature gradients on its equatorward side, where the meridional pressure gradients (not shown) and westerlies weaken substantially (Figure 2c). While this mechanism shares similarities with the polar amplification hypothesis in its reliance on meridional temperature gradients, it differs in its focus on land-driven warming as the primary driver, rather than amplified polar warming missing in the NH summertime simulations (Figure 2c). According to the thermal wind relationship, changes in the meridional temperature gradient are proportional to changes in the vertical wind shear. Figure 2f shows that the changes in the vertical wind shear are significantly correlated with the changes in the near-surface wind speed within and across individual seasons. Although other factors can contribute to wind changes (e.g., NH wintertime static stability in Figure 2e), the correlations in Figure 2f suggest a strong link between amplified land warming and wind stilling in the NH summertime.

**3.2 Emergence and Robustness of Regional Wind Changes**

The analyses of the zonal mean wind changes (Figure 2f) suggest that the signal of human-caused climate change (i.e., ensemble mean) can exceed the background noise of natural climate variability (i.e., ensemble spread) by the end of the 21$^{st}$ century. During the NH summertime, the ensemble mean changes are larger than the ensemble spread in some midlatitude regions (Figure 2f), indicating that human-forced changes in near-surface wind may become distinguishable from natural variability within this century. To understand when and where these changes may appear, we conduct a more granular analysis using the reference regions by the Intergovernmental Panel on Climate Change (IPCC) (Figure 3).

While the projected changes in average monthly near-surface wind speed are relatively small (<5%) during 2021–2050, they can reach up to 10% of the historical mean by 2071–2100. The most robust and consistent changes appear during the summertime in Europe and North America, where significant wind stilling is projected. The projected changes in the 30-year climate also show substantial natural variability, as indicated by the ensemble spread of relative changes. Notably, the natural variability in the regional mean changes (e.g., ~10% range in Figure 3e) is larger than that of the midlatitude zonal mean changes (~2%) (Figure 2f). This suggests a relatively low signal-to-noise ratio for detecting and attributing human-forced changes on the regional scale.

Due to the large natural variability, most regional averages do not exhibit changes that can be confidently attributed to human influences until after 2050 (Figure 3m). The earliest detectable wind signals are the projected wind increases in Africa, which may benefit the local wind development. Towards the end of the 21$^{st}$ century, summertime wind stilling signals emerge in North America, Europe, and East Asia. While the analysis of IPCC regions provides a valuable overview, it is important to recognize that averaging over large regions masks finer-scale variations. For instance, Fig. S6 suggests that detectable signals may emerge earlier in parts of the US West, East Canada, and Southern China. Such early wind stilling, if realized, could pose challenges for wind energy projects that start operation in the coming decades.

While the precise emergence timing of these warming-induced changes is subject to the specific analysis methods used (e.g., trend and detectability definitions) and potential biases in the CESM2 simulations, the findings suggest that the impacts of climate change on wind projects can emerge within the 21$^{st}$ century in high-emission scenarios. Additional analysis suggests that the projected wind stilling can be substantial but will unlikely undermine the cost-competitiveness of wind energy relative to fossil-fueled electricity generation, especially at sites with high-grade wind resources (Supplementary Materials). Therefore, the projected wind stilling will unlikely prevent wind energy from playing an increasingly important role in electricity generation.

**3.3 Potential of Compounding Energy Risks**

As wind energy development continues, the closely linked wind stilling and land warming may have increasingly important implications for electricity grid reliability in high-emission scenarios. To meet consumer needs and ensure equipment safety, the supply of electricity must be continuously adjusted to match the fluctuating consumer demand. Imbalances between supply and demand can lead to infrastructure damage, volatile energy prices, or catastrophic power outages. As global warming raises the summertime cooling demand for air conditioning and refrigeration, meeting this increasing cooling demand will need more renewable energy to mitigate further warming and reduce reliance on fossil fuels. Ensuring electricity resource adequacy in the NH midlatitudes could become more challenging if the projected summertime land warming and wind stilling occur.

The CESM2 simulations suggest substantial increases in energy demand for cooling during the NH summertime, particularly in Europe and North America (Figure 4a). Specifically, the cooling demand in the peak summer months, as measured by CDDs, can increase by 100–200 CDDs in 2071–2100 (Figure 4b–4d). The cooling demand also increases in the transition seasons, suggesting summer cooling periods will extend longer. The largest relative increases in CDDs appear in the mid-latitudes, exceeding 100% in much of Europe and North America (Figure 4a; Figs. S9 and S10). Even considering the relatively close Year 2050, substantial CDD increases (>50%) can occur in Europe (Figure 4c). These substantial increases can be attributed to the amplified land warming (Figure 1c) and the nonlinear definition of CDDs that make the metric particularly sensitive to the warming near the midlatitude temperature. The notable CDD increases in regions that historically had small cooling demand (e.g., parts of Europe) are consistent with recent research (e.g., Miranda *et al* 2023, Staffell *et al* 2023) and highlights the potential of rapid shifts in energy consumption patterns as the climate warms.

In high-emission scenarios, efforts to meet the surging summertime cooling demand can be complicated by the projected decline in wind energy output. The summertime wind stilling projected by the CESM2 can reduce wind energy output by as much as 25% in 2071–2100 in North America and Europe (Figure 4b–4d). When accounting for natural climate variability, the projected decline in wind energy output could reach 30–40% in some simulations (Figure 4a). While the comparison between CESM2 and other climate simulations suggest some differences in the magnitude and seasonality of wind speed changes (Supplementary Materials), the overall signal of reduced summertime potential of wind energy is consistent. This possible reduction in electricity generation, coupled with the increasing cooling demand, may raise concerns about the ability of electricity grids to maintain reliable power supply during the summer months, particularly in regions that are already facing challenges with resource adequacy (North American Electric Reliability Corporation 2023). Meeting these challenges may become increasingly hard if the amplified summer land warming and wind stilling materialize in the upcoming decades.

## 4. DISCUSSION

Building on past studies of near-surface wind changes, we use large ensemble climate simulations to identify robust changes and quantify uncertainties. Specifically, the large ensemble affords the opportunity to assess the sensitivity of near-surface wind changes to the large-scale environment changes, which helps highlight the contribution of amplified land warming to the summertime wind stilling. The large ensemble also helps estimate the emergence time of significant, human-forced wind stilling, which can be valuable for guiding wind project investments amidst a changing climate. Importantly, we hope that the identification of the intrinsic link between the amplified land warming and the summertime wind stilling will motivate future theoretical and applied research on climate-energy compound risks.

To contextualize the importance of compound risk of wind stilling and land warming, we emphasize that even seemingly small disruptions in energy supply and demand balance can have significant societal consequences. For example, less than 20% of the global natural gas supply was affected by the Russian invasion of Ukraine, yet the resulting supply-demand disequilibrium triggered a surge in global energy prices and pushed hundreds of millions of people into energy poverty (Guan *et al* 2023). In high-emission scenarios, the projected increases in cooling demand and decreases in wind energy output could eventually create similar or even larger imbalances without proactive planning and adaptation. To compensate these changes, additional investments in electricity generation capacity are necessary and may make the target of net zero emissions more costly to accomplish.

## 5. CONCLUSION

As the world increasingly relies on wind energy to decarbonize electricity generation, understanding the potential impacts of climate change on wind resources becomes crucial. By leveraging ~200 climate simulations, we highlight the large seasonal variations in the response of the near-surface wind speed to future anthropogenic warming. While the NH wintertime and SH changes are mainly associated with a poleward shift of the midlatitude westerlies, the summertime NH changes show widespread and pronounced wind stilling. The human-forced stilling signals can emerge within this century and are accompanied by amplified land warming. In a high-emission scenario (SSP3-7.0), the projected summertime speed decrease may reach up to ~10% of the historical climate wind speed, which could lead to a ~25% loss in potential wind energy output. The physical mechanism of this summertime wind stilling, while explored by some previous studies, has not been fully explained. We advance the physical understanding of the summertime wind stilling by demonstrating inconsistencies between existing interpretations and the CESM2 simulations; and proposing a new mechanism that emphasizes the amplified land warming and associated tropospheric temperature changes.

Importantly, our analysis highlights an under-appreciated risk of compounding climate and energy challenges that could lead to summertime energy shortages in the densely populated midlatitudes. This energy perspective complements earlier studies that explore the compounding of physical hazards (Zscheischler *et al* 2018), such as the co-occurrence of wind and rainfall extremes (Manning *et al* 2024). The risk highlighted by this study arises from the convergence of two key factors: a surge in energy demand for cooling due to rising temperatures and a warming-induced decline in wind energy output. The CESM2 simulations suggest that this compounding risk could lead to significant challenges for meeting energy demand during the summer months. In a high-emission scenario (SSP3-7.0), the populated mid-latitudes may experience up to ~140% increases in cooling demand and ~25% less energy output from wind turbines during the summer months of 2071–2100 compared to 1981–2010. To avert the seasonally amplified climate-energy risk, pairing wind energy with efficiency improvements, other clean energy (e.g., solar energy) (e.g., Tong *et al* 2021, Gernaat *et al* 2021), and grid infrastructure upgrades (e.g., Moraski *et al* 2023) is likely valuable for future energy investments.

**ACKNOWLEDGMENTS**

The author thanks Drs. Andrew Dessler, Flavio Lehner, Tiffany Shaw, and Yi Zhang for helpful discussions. The author thanks the data access and computing support provided by the National Science Foundation National Center for Atmospheric Research's Coupled Model Intercomparison Project (CMIP) analysis platform (doi:10.5065/D60R9MSP). G.Z. is supported by the United States National Science Foundation (Award 2327959) and the faculty development funds provided by the University of Illinois at Urbana-Champaign and the Gies Business School.

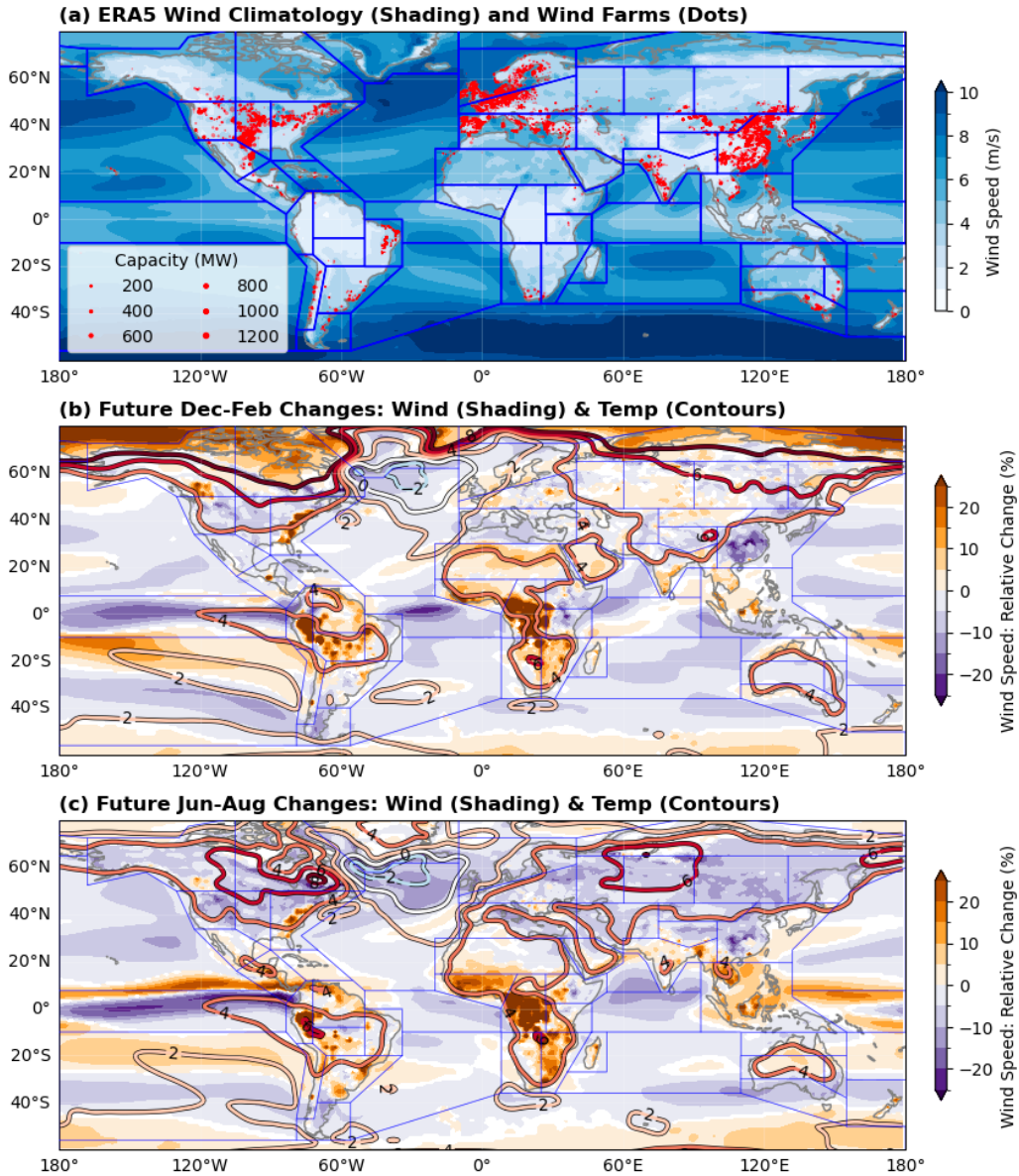

*Figure 1 Wind farm locations and future changes in surface temperature and near-surface wind speed. (a) The annual mean 10-meter wind speed (blue shading; m s$^{-1}$) from the ERA5 dataset (Methods) and the wind farm information (red dots) from the Global Energy Monitor. The ERA5 annual mean is calculated using the monthly data of 1981–2010. The size of the dots shows the capacity of individual wind farms. (b) Absolute changes in surface temperature (contours; K) and relative changes in 10-m wind speed (shading; %) of the December–February climate. The changes are the differences between the annual means of 2071–2100 and 1981–2010 simulated by*

*the CESM2 large ensemble (SSP3-7.0; Methods). The changes below the 99% confidence level (Student's t-test) are close to zero and masked out. (c) Same as (b), but for June–August. For visual clarity, the temperature changes are smoothed. Evaluating the relative changes in wind speed helps leverage the empirical relationship that suggests wind speed at different elevations experience the same relative changes (Methods). The blue polygons follow the Intergovernmental Panel on Climate Change (IPCC) climate reference regions* (Iturbide *et al* 2020).

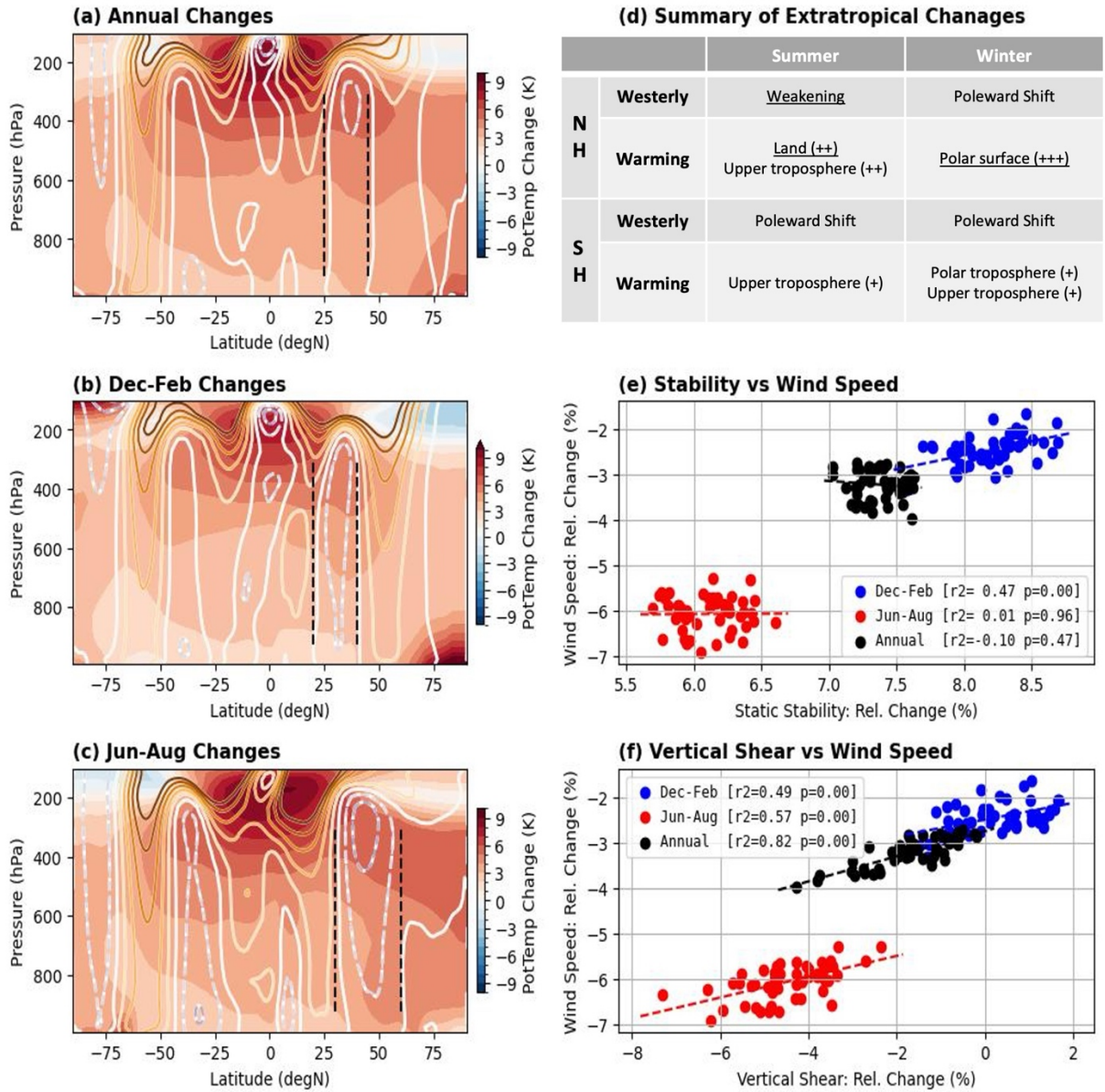

*Figure 2 Changes in the zonal averages of wind, temperature, and related variables in the CESM2 simulations (SSP3-7.0). (a) Ensemble mean changes in the annual means between 2071–2100 and 1981–2010. The color shading represents changes in the potential temperature (K). The contours show changes in the zonal wind (m s$^{-1}$) in a coordinate of latitude and vertical pressure, and the contour interval of 0.25 m s$^{-1}$. For clarity, the negative values are denoted with dashed contours, and the values with magnitudes greater than 2.5 m s$^{-1}$ are skipped. The negative values (dashed*

*lines) indicate a weakening of westerly winds. The black vertical dashed lines highlight where westerlies weaken in the northern midlatitudes. (b) Same as (a), but for the December–February means. (c) Same as (a), but for the June–August means. (d) Summary of extratropical wind and temperature changes in the summertime and wintertime of both hemispheres. The number of plus signs indicates the warming magnitudes qualitatively. (e) The relationship between the relative changes in the near-surface wind speed and static stability (300–925 hPa) between the 2071–2100 and 1981–2010 climate. The analyses were conducted for the December–February, June–August, and annual means of each ensemble member (dots). The analyzed regions are between the vertical dash lines in a–c, where midlatitude westerlies weaken. The dashed lines in (e) show the linear regression lines with their r-square and p-value denoted in the legend. (f) Same as (e), but for the vertical wind shear (300–850 hPa).*

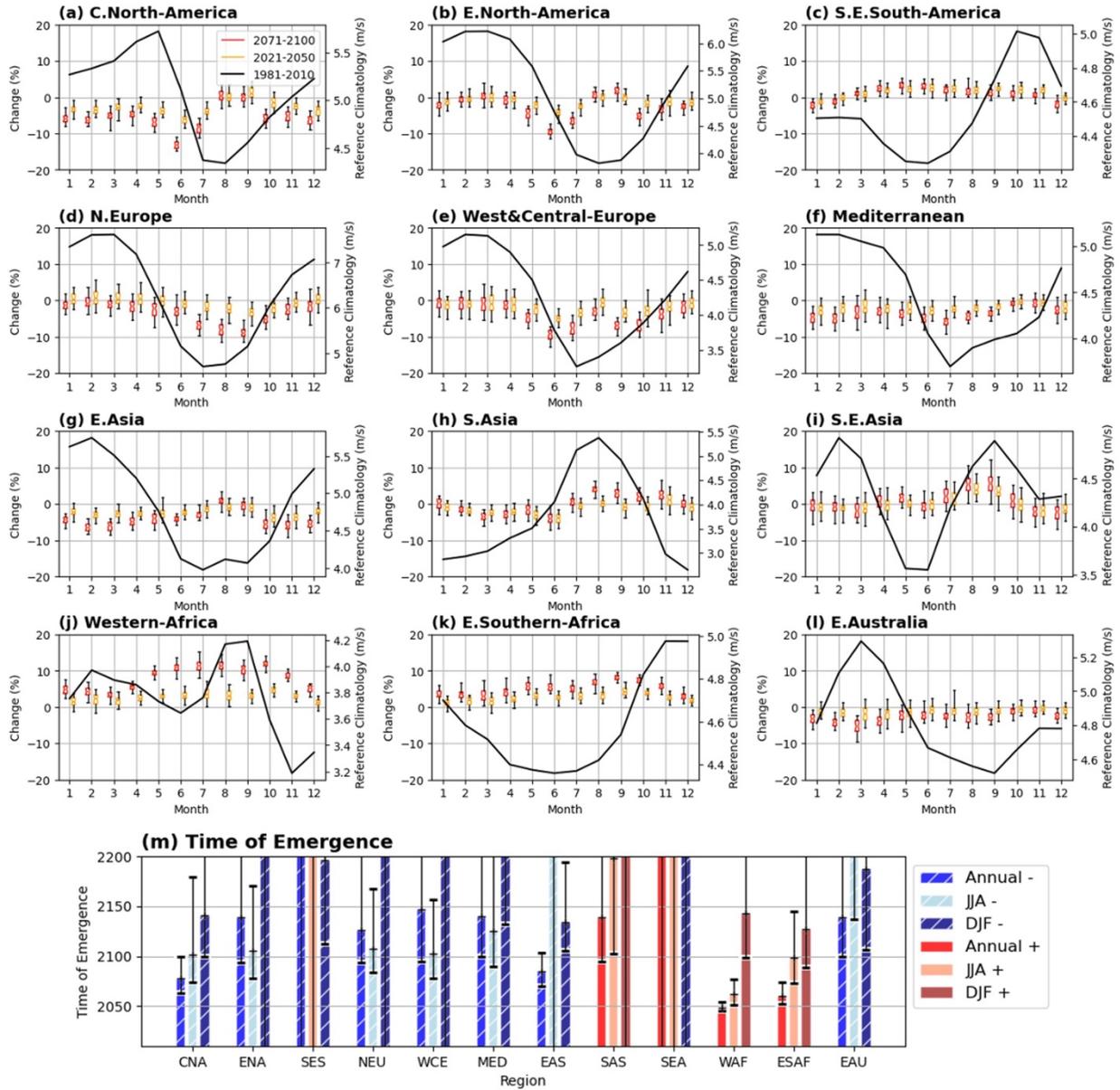

*Figure 3 Future changes in regional wind speed and the time of emergence for statistically significant changes in CESM2 simulations (SSP3-7.0). (a) Relative changes of near-surface wind speed in Central North America (CNA). The ensemble statistics of 2021–2050 (yellow) and 2071–2100 (red) are indicated with boxplots. The boxplots show 0.5th, 25th, 50th, 75th, and 99.5th percentiles. The absolute values of the CESM2 ensemble mean historical climate (1981-2010) are represented with the black line and scale with the axis on the right. (b–l) Same as (a), but for*

*Eastern North America (ENA), Southeastern South America (SES), North Europe (NEU), West and Central Europe (WCE), Mediterranean (MED), East Asia (EAS), South Asia (SAS), Southeast Asian (SEA), Western Africa (WAF), Eastern Southern Africa (ESAF), and East Australia (EAU). The changes in regional averages are calculated based on selected climate zones in Figure 1. (m) Time of emergence (TOE; Methods) for the ensemble mean changes in the CESM2 large ensemble simulations. The increase of wind speed is denoted with warm colors, while the decrease of wind speed is denoted with cold colors and hatching patterns. The error bars denote the absolute value range of 2.5$^{th}$ and 97.5$^{th}$ percentiles of ensemble members. The estimate is conducted in the same regions as in Figure 3a–3l for the annual, Jun-Aug (JJA), and Dec-Feb (DJF) means of the near-surface wind speed.*

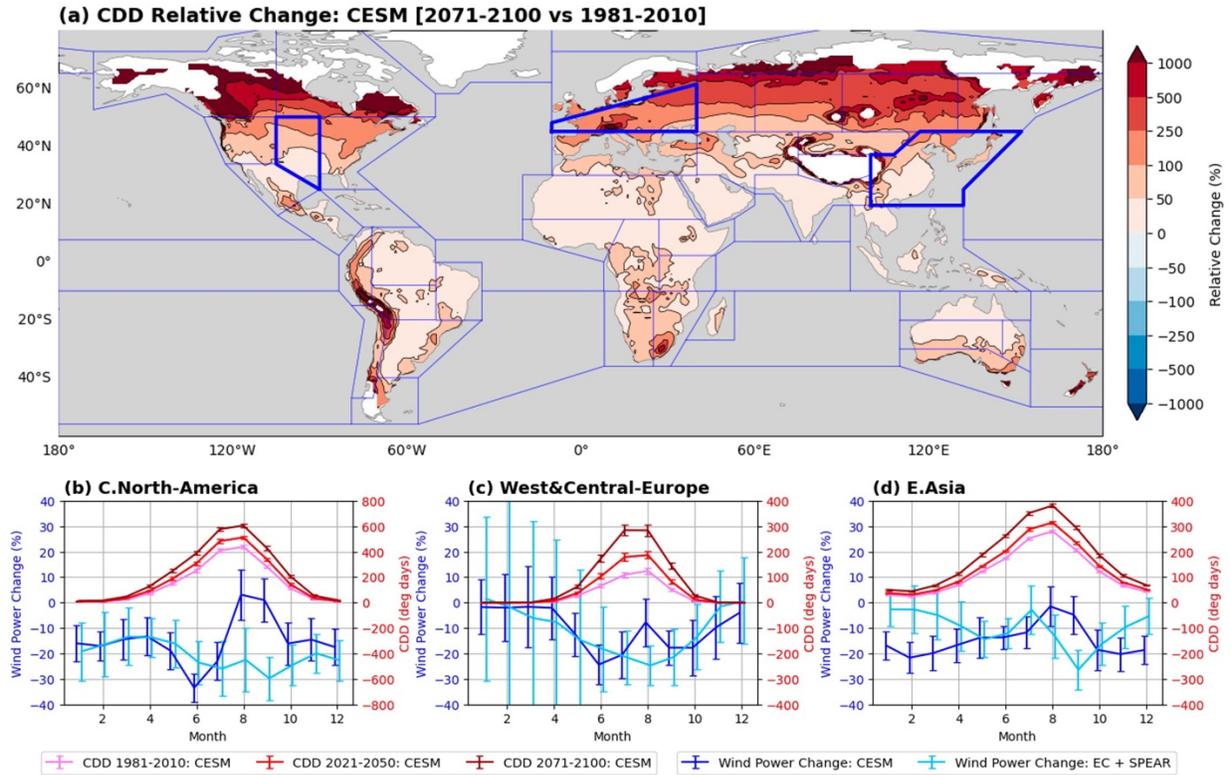

*Figure 4 Future changes in the cooling degree day (CDD) and the near-surface wind speed. (a) Relative changes in the annual mean CDD (ºC days) during 2071–2100 relative to 1981–2010. The land regions with little cooling demands (CDD<100 ºC days) and the oceanic regions have small population sizes and are masked out for figure clarity. The blue polygons in (a) follow Figure 1. (b) Monthly CDD climate and the relative changes of wind power output in Central North America. The CDDs are calculated for three simulation periods (1981–2010, 2021–2050, and 2071–2100) using the CESM2 large ensemble data (SSP3-7.0) and $T_{ref}$=18ºC. The wind power estimates (Methods) from the CESM2 large ensemble (blue) and the SPEAR (SSP5-8.5) and EC-Earth (SSP3-7.0 and SSP5-8.5) large ensembles (cyan) are denoted separately. The relative changes are evaluated by comparing 2071–2100 to 1981–2010. The error bars indicate the ranges of two standard deviations. (c) Same as (b), but for West and Central Europe. (d) Same as (b), but for East Asia. The three regions in (b–d) are highlighted with thick lines in (a).*

Supplementary Materials for

**Amplified Summer Wind Stilling and Land Warming Compound Energy Risks in Northern Midlatitudes**

Gan Zhang[1*]

[1]Department of Climate, Meteorology, and Atmospheric Sciences, University of Illinois at Urbana-Champaign, 1301 W. Green Street, Urbana, IL 61801

*Corresponding Author: Gan Zhang (gzhang13@illinois.edu)

This file includes:

Figures S1 to S11

Table S1

**Supplementary Description of Data**

This study uses the CESM2 large ensemble because of practical considerations related to data accessibility and processing. While the high resolutions of SPEAR and EC-Earth help simulate wind extremes, they also greatly increase the burden of data transferring and processing. For example, calculating daily degree days for the SPEAR or EC-Earth ensemble will need to transfer >10 TB of data. We estimate this task alone will involve a data volume comparable to or greater than the past CMIP5 and CMIP6 studies of future wind changes (Karnauskas *et al* 2018, Zha *et al* 2021, Deng *et al* 2022). In comparison, all the CESM2 data, including high-frequency outputs, are accessible via a file system at the US National Center for Atmospheric Research (NCAR). Finally, CESM2 is a community project that shares components with many other CMIP6 models, so the findings based on CESM2 might be leveraged by more model developers and users in the future.

The analyzed CMIP6 models have relatively high resolutions (100–200 km grid spacing) compared to the other CMIP6 models. The twelve models include GFDL-CM4, TAI-ESM1, BCC-CSM2-MR, CMCC-CM2-SR5, CSIRO-ACCESS-CM2, EC-Earth3, MIROC6, HadGEM3-GC31-MM, MPI-ESM1-2-HR, MRI-ESM2-0, NCAR-CESM2, NCC-NorESM2-MM. Details of the individual models are available in the associated documentation pages of the Earth System Grid Federation (ESGF).

The analysis in Supplementary Information used the historical population data (Center For International Earth Science Information Network-CIESIN-Columbia University 2018) from the Socioeconomic Data and Applications Center of the US National Aeronautics and Space Administration (NASA), as well as the historical degree day data (Mistry 2019) calculated using the data from the Global Land Data Assimilation System (Rodell *et al* 2004) of NASA. The analysis of project cost-competitiveness assumed the range of interest rates by consulting the US bank prime loan rate (1991–2023) published by the Federal Reserve Bank of St. Louis.

**Shared Socioeconomic Pathways**

The global climate simulations of the large ensemble simulations and CMIP6 are conducted with pre-determined forcings outlined by Shared Socioeconomic Pathways (SSPs) (O'Neill et al. 2017). The labels SSP3-7.0 and SSP5-8.5 denote the development paths of society (SSP3 or SSP5) and the additional radiative forcing by the year 2100 (7.0 W m$^{-2}$ and 8.5 W m$^{-2}$). SSP3 and SSP5 correspond to regional rivalry and fossil fueled development scenarios, respectively. The SSP3-7.0 corresponds to the upper-middle part of the radiative forcing range described by previous climate studies, while and SSP5-8.5 are at the upper bounds of radiative forcing. Assuming climate system responses scale with the forcing strengths, the simulations with SSP3-7.0 and SSP5-8.5 help highlight signals associated with anthropogenic warming. Therefore, the choice of analyzing SSP3-7.0 and SSP5-8.5 simulations does not indicate they are treated as the most likely scenarios.

**Estimates of Turbine-level Wind Speed and Energy Output**

Few GCM simulations output wind data at the turbine elevation (~100 m), but most save the 10-meter wind speed data. Assuming the wind speed in the planetary boundary layer follows the power law, the turbine-level wind can be estimated using an empirical wind-height relationship:

$$U_z = U_{ref} \left(\frac{H_z}{H_{ref}}\right)^\alpha \quad (1)$$

Where *U* stands for wind speed and *H* stands for measurement height. The subscript *ref* indicates a reference height (e.g., 10 m), and the subscript *z* indicates the level of interest (e.g., turbine level). The parameter $\alpha$ is typically a value around 0.11 but will be cancelled out in the ensuing analysis.

For the large ensemble simulations, extrapolating the 10-meter wind speed to turbine levels is computationally intensive. The resolution of climate simulations used in this study (~50 to 100 km) is unlikely adequate to accurately model terrain drags and the associated vertical wind shear. But if one assumes $(H_z/H_{ref})^\alpha$ in constant in the historical and future climates, the scaling in Eq. (1) suggests relative changes (e.g., late-century versus historical) would hold the same values at the 10-meter level and the turbine level. By circumventing the need of extrapolating wind speed, assessing relative changes greatly alleviate the burden of computation. Additionally, assessing relative changes also makes it straightforward to compare projections by various GCMs. In contrast, comparing absolute changes across models would need additional bias corrections that are computationally demanding and can complicate interpretations.

The mathematical derivations that link the relative changes at 10-meter and turbine-level wind speed are as follows. Starting with Eq. (1), one may write the wind speed in the historical and future climates as:

$$U_{z0} = U_{ref0} \left(\frac{H_z}{H_{ref}}\right)^{\alpha 0} \qquad (2)$$

$$U_{z1} = U_{ref1} \left(\frac{H_z}{H_{ref}}\right)^{\alpha 1} \qquad (3)$$

The subscript 0 and 1 indicates the historical and future climates, respectively. The subscript ref and z indicate a reference height (e.g., 10 m) and a turbine-level height (e.g., 100 m). Assuming no land use changes (e.g., urbanization), the coefficients $\alpha 0$ and $\alpha 1$ can be assumed to be approximately the same. Use this approximation and let $\beta \equiv \left(\frac{H_z}{H_{ref}}\right)^{\alpha 0}$, the two equations above can be rewritten as:

$$U_{z0} = U_{ref0}\, \beta \qquad (4)$$

$$U_{z1} = U_{ref1}\, \beta \qquad (5)$$

Now consider the relative changes of future climate at the reference and turbine-level heights:

$$\delta_{ref} = (U_{ref1} - U_{ref0})/U_{ref0} \quad (6)$$

$$\delta_z = (U_{z1} - U_{z0})/U_{z0} \quad (7)$$

Plugging Eqs. (4) and (5) into (7) and eliminating the coefficient $\beta$ yields:

$$\delta_z = (U_{ref1} - U_{ref0})/U_{ref0} \quad (8)$$

A comparison of Eqs. (6) and (8) suggests the relative changes at the reference level and the turbine level are equal. This serves as the basis for using relative changes in 10-m wind to approximate the relative changes in turbine-level wind. To estimate absolute changes at an arbitrary wind turbine level, one can consider Eq. (8) and use the following expression:

$$U_{z1} - U_{z0} = \delta_z U_{z0} = [(U_{ref1} - U_{ref0})/U_{ref0}] U_{z0} \quad (9)$$

This expression does not involve any vertical interpolation. It can work for arbitrary heights where Eq. (1) and the other pre-stated assumptions hold. When $U_{z0}$ is known, this expression can help estimate future wind speed under certain assumptions of relative changes (e.g., -5% and -10%). This method is applied to modify the observed turbine-level wind speed at wind farm sites, which serve as input for the latter analytics ("Modeling of Wind Farm Cost-Competitiveness").

While the study mainly analyzes wind speed changes, we also estimate potential changes in wind power output when discussing the risk of summer energy shortage (Figure 4). Our estimate of wind power output uses the monthly mean wind and thus includes biases. Specifically, the wind speed used in this study is calculated at model steps but averaged and saved at monthly time steps. When assessing the power output, calculating the cube of monthly means of wind speed differs from calculating the cube of wind speed at model steps and averaging on the monthly scale. This approximation is a practical compromise related to the lack of high-frequency wind speed data and follows Karnauskas et al. (2018). The same study also discussed the limitations of this approximation in detail.

**Electricity Grid Modeling and Degree Days**

Modeling the response of electricity grids to future climate change is a complex task that requires making assumptions about future infrastructure development, technological advancements, and socioeconomic changes (Craig *et al* 2022). These assumptions about these non-climate factors such as the rate of transportation electrification are often highly uncertain. Past experience suggests that such modeling methods can also systematically underestimate the growth of renewable energy (Creutzig *et al* 2017). Even if precise quantitative predictions are difficult to achieve at this stage, we argue that these challenges should not prevent us from proactively exploring the potential implications of climate change for electricity grids. To simplify the analysis and focus on the climate drivers of grid stress, we use cooling degree days (CDDs) as a proxy for energy consumption associated with indoor cooling. CDDs are a widely used metric that reflects the temperature difference between the daily average temperature and a reference temperature (e.g., 18°C), providing an indication of the need for cooling. Using different reference temperature values does not alter our findings qualitatively (cf. Figure 4 and Figs. S9 and S10).

**Inconsistencies Between CESM2 Simulations and Pre-existing Physical Interpretations**

This section extends the discussion of Figure 2 in Section 3.1 and highlights the inconsistencies between the CESM2 simulations and several pre-existing hypotheses that sought to explain the decrease of near-surface wind speed simulated by climate models.

*Polar amplification hypothesis.* In the CESM2 simulations, the polar surface warming in the NH is strong during the wintertime (Figure 2b) but is absent during the summertime (Figure 2c). If the polar amplification contributes the simulated summertime wind stilling, the amplified warming signal should be present in the summertime to meaningfully modulate the meridional temperature gradients.

*Surface roughness hypothesis.* The simulated weakening of midlatitude westerlies is widespread, including regions that lack significant land vegetation (Figure 1c). If vegetation changes were the dominant driver, the weakening would likely be more variable around the coast (e.g., the British Isles) and weaker over the ocean upstream (e.g., North Atlantic). Furthermore, this hypothesis cannot explain why the upper troposphere wind weakens more than the lower troposphere (Figure 2c).

*Stability-mixing hypothesis.* While static stability does increase in the region of summertime wind stilling (Figure 2c), an examination across the individual ensemble members suggests the stability changes lack a significant correlation with surface wind changes (Figure 2e). In fact, when considering the summertime, wintertime, and annual changes together, the ensemble members suggest that relatively large increases in static stability are associated with less wind weakening, contradicting the relationship suggested by this stability-mixing hypothesis.

**Example of Estimating Time of Emergence**

This section uses example data to provide a detailed technical description of the Time of Emergence (ToE) calculation. The example uses the annual mean wind averaged in the climate zone of North Europe (Fig. S5), where a decrease of wind speed is projected by the CESM2. The examined simulation years range from 1981 to 2100, and the period 1981–2010 is considered as the modern reference climate. We first focus on the reference period and calculate the ensemble mean of climate statistics including the 30-year climate mean and the standard deviations of individual ensemble members. These values are used to characterize the mean state and natural variability of the reference climate. The next step is to focus on the period of 2011–2100 and use the linear regression to estimate the long-term climate trend for the ensemble mean. The regression helps estimate the time when the ensemble mean may intersect with the two-standard-deviation threshold, which would indicate a long-term trend has shifted the climate to the extent that it substantially differs from the reference period. In Fig. S5, this intersection occurs around Year 2125 and 115 years after 2010. Year 2125 is considered as the ToE for the climate change signals to emerge from natural variability. The same regression is conducted for individual ensemble members to estimate the uncertainty of the ToE. The lower bound of the error bars use the absolute value of the difference between $2.5^{th}$ and $50^{th}$ percentile values, while the upper bound of the error bars use the absolute value of the difference between $50^{th}$ and $97.5^{th}$ percentile values.

**Comparison with SPEAR and EC-Earth Simulations**

To assess the robustness of the regional wind changes simulated by CESM2, particularly the summertime wind stilling in the NH midlatitudes, we examine large-ensemble simulations from two other high-resolution GCMs (see Methods; Figs. S7–*S*8). The inclusion of these additional models, along with their large ensemble sizes (>100 members), provides a broader range of potential climate outcomes and helps to delineate the uncertainty associated with the projections. For example, the GCMs suggest decreases in wind speed in East Asia and the Mediterranean but disagree on the seasonal timing of these potential decreases. Furthermore, the three models show disagreement regarding the direction of future wind changes in parts of Africa,

South Asia, and Southeast Asia. This suggests that the projected increase in wind resources reported by earlier studies (Karnauskas *et al* 2018, Zha *et al* 2021) is subject to considerable regional uncertainty. Despite these disagreements, the simulations consistently highlight the summertime wind stilling in Europe and North America. Due in part to stronger $CO_2$ forcing in some experiments (SSP5-8.5), the projected reductions in wind speed can reach 10–15%, exceeding those simulated by CESM2 (SSP3-7.0) (Figs. S7–*S*8). Furthermore, a comparison of the changes in 2021–2050 (Fig. S7) and 2071–2100 (Fig. S8) relative to the reference climate (1981–2010) suggests the magnitude of wind speed changes is larger in the late century, when the greenhouse gas concentration is more elevated. These findings suggests that the magnitude of the wind speed decrease scales with the level of future greenhouse gas emissions.

**Modeling of Wind Farm Cost-Competitiveness**

This section and the ensuing financial analyses seek to illustrate potential impacts of the projected decrease of near-surface wind speed on wind energy investments. An important driver of wind farm investments is their cost-competitiveness relative to alternative energy sources. The cost-competitiveness of wind farms depends on project costs and incentives that vary across countries. For simplicity, our assessment focuses on the US wind projects and uses the System Advisor Model (SAM, version 2022.11.21) developed by the US National Renewable Energy Laboratory (NREL). The SAM is a free model that provides information on wind turbine parameters, wind resource data, reference financial assumptions, and the US policy incentives.

When evaluating the energy output of wind farms, we use the GE 2.5 MW XL model as the reference wind turbine and set the hub height to 80 m. The LCOE in the reference scenario (low interest and no wind change) is approximately 0.045 USD per kWh, near the middle of the unsubsidized LCOE of onshore wind projects in 2023 (Bilicic and Scroggins 2023). While newer wind turbine models may be considered, the project costs are harder to estimate due to data latency and conflicting price drivers (e.g., recent inflation and efficiency increase). We anticipate this GE model choice makes the LCOE estimates too conservative (which is reasonable for stress testing), especially if inflation is abated in the future.

The analysis of wind farm cost-competitiveness uses the hourly wind data provided by the NREL of the US Department of Energy. The data was built with twelve selected months from a multi-year period that describes weather conditions during 1997–2010. The dataset consists of wind information at turbine operation elevation (e.g., 80 m) and is intended for feasibility and policy studies of wind energy projects. The proprietary data shared by NREL is at representative, anonymized wind farm sites across the US. The precise location was masked out by NREL to protect the interest of the data providers. We adopt the representative wind farm sites provided by SAM but exclude the offshore sites. Despite their potentials, offshore wind farms currently only account for a small portion of existing wind farms in the US and have much higher construction costs than their onshore counterparts. The remaining 28 representative sites cover a wide range of terrains including flat lands, canyon lands, rolling hills, and mountainous regions. Additional details of these wind farm sites are available in Table S1.

When estimating potential cost-competitiveness changes related to future wind stilling, we modify the reference hourly wind data by multiplying a fixed factor (e.g., 0.95) but keep the other physical data (e.g., wind direction and humidity) unchanged. This approach leverages the findings about relative changes in wind speed ("Estimates of Turbine-level Wind Speed and Energy Output") and serves to illustrate the cost-competitiveness sensitivity and estimate the bounds of potential losses. To compare the cost-competitiveness of wind projects, we consider two financial metrics, the levelized cost of energy (LCOE) and the internal rate of return (IRR). The LCOE is denoted in cents per kilowatt-hour of electricity and represents the cost of electricity over the project lifecycle. The IRR is the rate of return when the net present value of future cash flows is set to be equal to the initial investment. The LCOE and the IRR are used to evaluate the financial competitiveness of wind energy projects. The lower the LCOE or the higher the IRR is, the more attractive a wind energy project is for investors. The SAM calculation of the LCOE and the IRR accounts for electricity sales, project costs (e.g., construction, operation, and financing), as well as tax liability and incentives. For example, the calculation considers the production tax credit at 0.026 USD per kWh that escalates 2.5% per year with a 10-year term. Our calculations use the SAM-recommended values that were based on the recent NREL cost analysis (Stehly and Duffy 2022). Technical details about how the model evaluates the LCOE and the IRR are available in the NREL user documentation (National Renewable Energy Laboratory n.d.).

Since estimating future engineering advancements and finance-policy environments is not straightforward and requires great efforts, we instead ask how the projected wind stilling might affect wind energy projects if such wind changes occur in the early 2020s. Our analysis considers a range of scenarios that cover the wind stilling projected by the large ensemble simulations (0–10% by 2100) and the historical prime loan rate (3.25–9.5% in 1991–2023). The prime loan rate is likely the most important financial parameter as the interest compounding can greatly amplify the other costs over the project lifetime. As an illustrative analysis, we present results of six scenarios (Fig. S10) and discuss the sensitivity of the LCOE and the IRR to potential wind changes. The environment with 4% interest rate of prime loans and no wind changes resembles 2020 and serves as the reference state. Replacing the interest rate with 8% leads to an environment that resembles 2023, when the cost-competitiveness of wind energy projects was challenged by rising financial costs. Interested readers can find the input parameters used in the cost-competitiveness analysis in the code repository and explore more scenarios.

**Cost-competitiveness of Wind Energy Projects**

The cost-competitiveness of renewable energy is a crucial factor driving investment decisions and accelerating the ongoing energy transition. To our knowledge, past academic studies that analyzed GCM-simulated wind stilling have not explicitly considered its implications for the cost-competitiveness of wind energy projects. While onshore wind energy has become one of the most affordable electricity sources(Bilicic and Scroggins 2023), the potential for future wind stilling adds a layer of complexity to investment decisions, as investors need to carefully evaluate investment options across different regions and energy sources (e.g., wind versus solar energy). For individual wind projects, a 10% reduction in near-surface wind speed may lead to a 20–30% reduction in electricity output, as the power generated by a wind turbine is proportional to the cube of the wind speed. Assuming no changes in the capital or operational costs of wind projects, this reduced electricity generation can erode the cost-competitiveness of a project or threaten its financial viability. Therefore, estimating the financial impacts of warming-induced wind stilling is essential for informed investor decision-making and for ensuring the long-term financial viability of wind energy projects.

To assess the potential financial implications of climate change for wind energy investments, we develop a set of scenario analyses to stress test the financial returns and electricity generation costs of wind projects under different climate and finance conditions. These analyses use the System Advisor Model (SAM) developed by the National Renewable Energy Laboratory (NREL) to simulate the financial performance of wind projects, considering factors such as the range of wind speed changes and interest rate changes. To simplify the cost-competitiveness analysis and to isolate the impact of climate and interest rate changes, we assume the equipment and engineering costs and policy incentives (e.g., production tax credit) remain at the level of the early 2020s. This enables a direct comparison of the financial risks associated with climate change and interest rate changes. For example, the rapid increase in the interest rates during 2022–2023 significantly increased the cost of commercial loans in the US, contributing to widespread delays or cancellations of less cost-competitive wind energy projects (Vakil and DiSavino 2024). We explore if warming-induced wind stilling may have similar or greater consequences on wind energy development. Specifically, we analyze six finance-climate scenarios by pairing two extreme interest rate scenarios (4% and 8%) with three potential wind change scenarios (no change, 5%-decrease, and 10%-decrease). To simplify the demonstrative analyses and leverage the open-source NREL model, which is tailored to the US context, we focus on 28 representative sites of wind farms in the US.

The scenario analyses reveal that the warming-induced wind stilling can substantially impact the cost-competitiveness of wind energy projects, particularly in terms of increasing the levelized cost of energy (LCOE) (Fig. S10). We use a reference scenario that approximates the 2020 environment with a 4% commercial loan rate and no wind change. If the near-surface wind speed decreases by 10% (Fig. S10a) relative to the baseline scenario, the LCOE can increase by 38% (equivalent to 0.017 USD / kWh) when averaged over all the representative sites. In comparison, the levelized costs of wind energy shows a smaller increase (<15%; equivalent to a change of 0.005 USD / kWh) even with a notable 4% increase in the interest rate that resemble the period of 2022–2023. Interestingly, the sensitivity to wind stilling at the high-LCOE sites appears higher than the low-LCOE ones, suggesting resilient advantages of the low-LCOE sites.

In the absence of subsidies, the potential cost increase associated with wind stilling is likely adequate to make the onshore wind energy less cost-competitive than the utility-scale solar energy (mean LCOE ~ 0.024 USD / kWh) (Bilicic and Scroggins 2023), but the onshore wind energy will likely remain cost-competitive compared to fossil fuel-based alternatives in the US (mean LCOE ~ 0.039 to 0.068 USD / kWh) (Bilicic and Scroggins 2023). Regarding the investment return rate (IRR), the most notable impact of wind stilling occurs in the low-interest environment, where a 10% decrease in wind speed reduces the average internal return rate from 11.6% to 2.8%. In these simplified scenarios, such a reduction can turn wind energy projects unprofitable at ~29% of the examined wind farm sites. This represents a substantial portion (33%) of the sites that would have been profitable under the low-interest, no-wind-change scenario (IRR+; Fig. S11c). While these demonstrative analyses provide valuable insights into the potential financial impacts of wind stilling, it is important to note that real-world investment decisions are influenced by a wider range of factors (e.g., subsidies and financing arrangements) that need to be accounted by more complex models.

To further explore the potential financial risks associated with wind stilling, we now focus on a stress test scenario that combines strong wind stilling (-10%) with a high interest rate (8%). This stress testing estimates the upbound of climate impacts that uses conservative cost parameters and ignores technology advancements (e.g., efficiency improvements). Comparing this adverse scenario with a favorable, 2020-like reference scenario (no wind changes and low interest rate) allows us to estimate the upper bound of the potential negative impacts of wind stilling. In this adverse scenario, the LCOE of wind projects increases substantially, though the LCOE remains below 0.05 USD / kWh at 25% of the examine sites (Fig. S11b) and thus stays cost-competitive relative to fossil fuel-based electricity. Unsurprisingly, the profitable sites also dropped substantially compared to the reference scenario (cf. Figures S11c and S11d). Nonetheless, the average internal return rate of the modelled wind farms in the US remain positive (0.6%) in this extreme scenario and still exceeds 5% at the most profitable sites (Fig. S11d). Therefore, wind energy investments in the US can likely stay profitable even with extreme wind stilling and adverse investment environments. Through careful site selection, advancing equipment engineering designs, and effective mitigation of greenhouse gas emissions, the cost-competitiveness of wind energy projects will likely stay attractive in the coming decades in the US.

While the estimates with the NREL model incorporate simplifying assumptions and are specific to the US context, we anticipate that the general conclusion – that wind energy can remain cost-competitive even under challenging climate and economic conditions – would hold true for global wind investments.

**Data Availability**

The data generated by this study are available to reviewers and will be deposited at Zenodo upon acceptance. The raw model data and observational data are available from third-party providers under research use licenses. The ERA5 reanalysis is available at the National Center for Atmospheric Research (NCAR) Research Data Archive (https://rda.ucar.edu/datasets/ds633-0/). The data of the CESM2 Large Ensemble Community Project, with supercomputing support by the IBS Center for Climate Physics in South Korea, is available via NCAR (https://www.cesm.ucar.edu/community-projects/lens2). The SPEAR large ensemble data is available via the US National Oceanic and Atmospheric Administration (NOAA) (https://www.gfdl.noaa.gov/spear_large_ensembles/). The EC-Earth large ensemble data and the other CMIP6 simulations are available from the Earth System Grid Federation (ESGF) of the Lawrence Livermore National Laboratory (https://esgf-node.llnl.gov/projects/esgf-llnl/).

The bank loan interest rate is accessible via the US Federal Reserve Bank of St. Louis (https://fred.stlouisfed.org/series/PRIME). The Global Energy Monitor provided the wind project data (https://globalenergymonitor.org/projects/global-wind-power-tracker/). The historical population data is from the Socioeconomic Data and Applications Center of the US NASA (https://sedac.ciesin.columbia.edu/data/collection/gpw-v4/population-estimation-service). The historical degree data used in Supplementary Information are shared by Mistry(Mistry 2019) (https://doi.pangaea.de/10.1594/PANGAEA.903123). The SAM model and its Python implementation (PySAM) are available from the NREL (https://sam.nrel.gov/).

The code used to generate Figures 1–4 is available at Zenodo (10.5281/zenodo.13922203) for public access. Additional analysis code and data are available from G.Z. upon request on a case-by-case basis.

**Supplementary References**

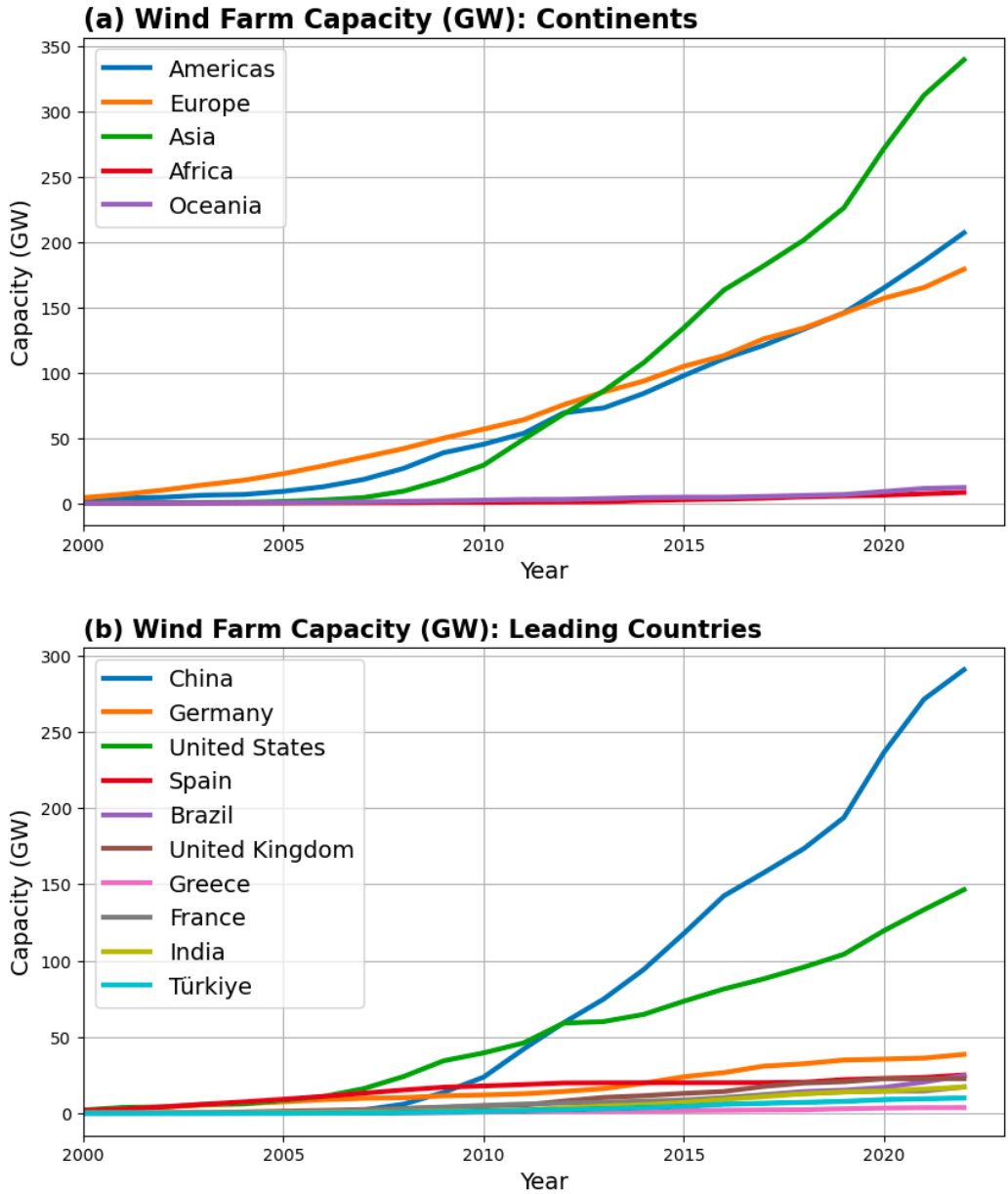

*Fig. S1 Growth of the installed capacity (GW) of wind farms during 2000–2022. (a) Capacity growth by continents. (b) Capacity growth of the ten counties with the highest installed capacity in 2022. Source data is provided by Global Energy Monitor.*

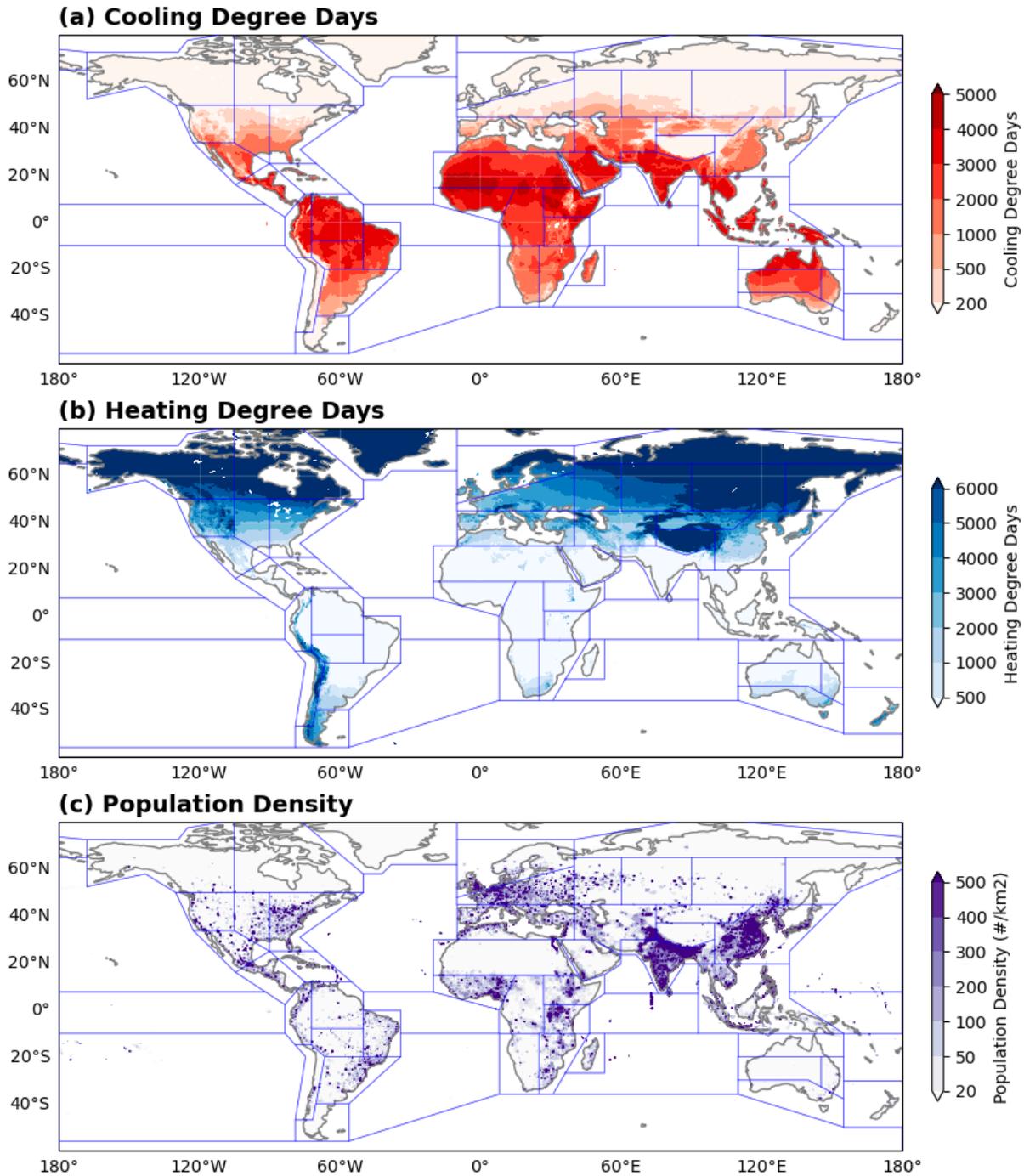

Fig. S2 Climate of the degree days (1981–2010) and the population density (2020). (a) Annual mean of cooling degree days (ºC day). (b) Annual mean of heating degree days (ºC day). (c) Population density (km$^{-2}$). The polygons indicate the climate zones (Iturbide *et al* 2020). Source data is from Mistry (Mistry 2019) and NASA's Socioeconomic Data and Applications Center.

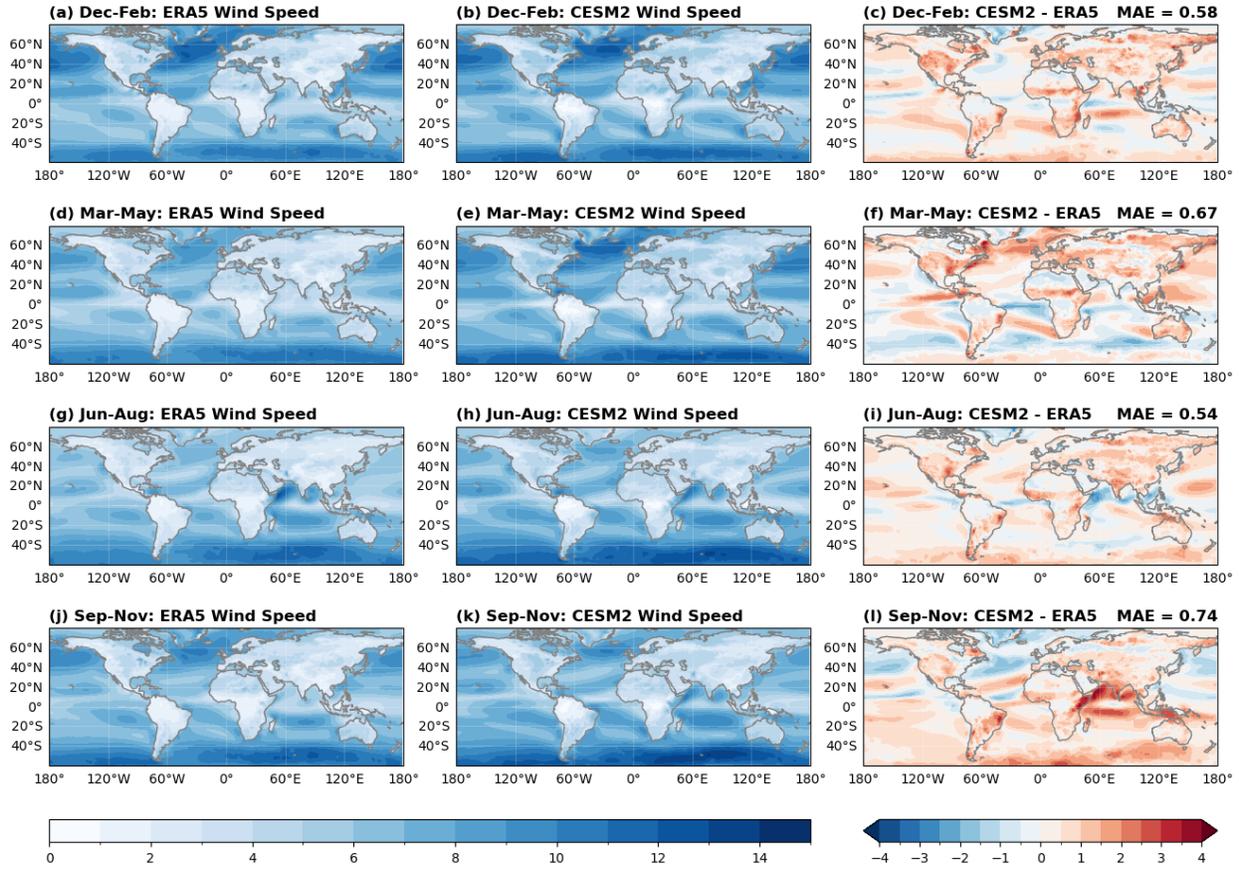

*Fig. S3 Near-surface wind speed from the ERA5 and the CESM2 large ensemble simulations. (a) ERA5 wind speed (10-m; m s$^{-1}$) climate during December–February of 1981–2010. (b) CESM2 wind speed (10-m; m s$^{-1}$) climate during December–February of 1981–2010. (c) The difference between the CESM2 and ERA5 climate. The mean absolute error (MAE) of the domain (60ºS–80ºN) is denoted in the upper right. (d)(g)(j) Same as (a), but for March–May, June–August, and September–November. (e)(h)(k) Same as (b), but for March–May, June–August, and September–November. (f)(i)(l) Same as (b), but for March–May, June–August, and September–November.*

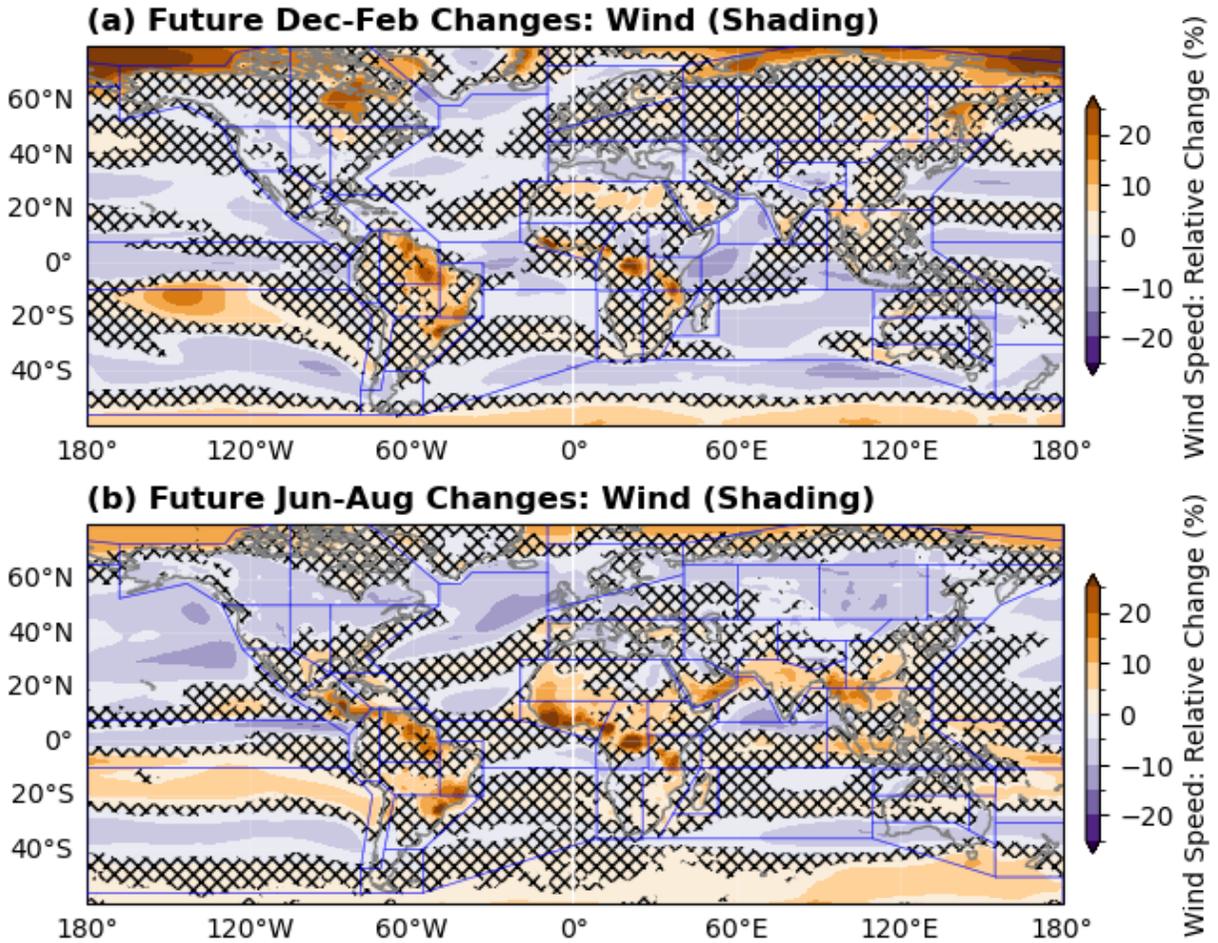

*Fig. S4 The surface wind changes projected by twelve CMIP6 models. The analysis periods are the same as Figure 1b and 1c, but the climate forcings are replaced with the SSP5-8.5 to highlight responses to warming with the relatively small CMIP6 ensemble. The plotted variable is in the ensemble average of relative changes in near-surface wind speed. The hatching indicates regions where fewer than 50% models agree on the sign of changes.*

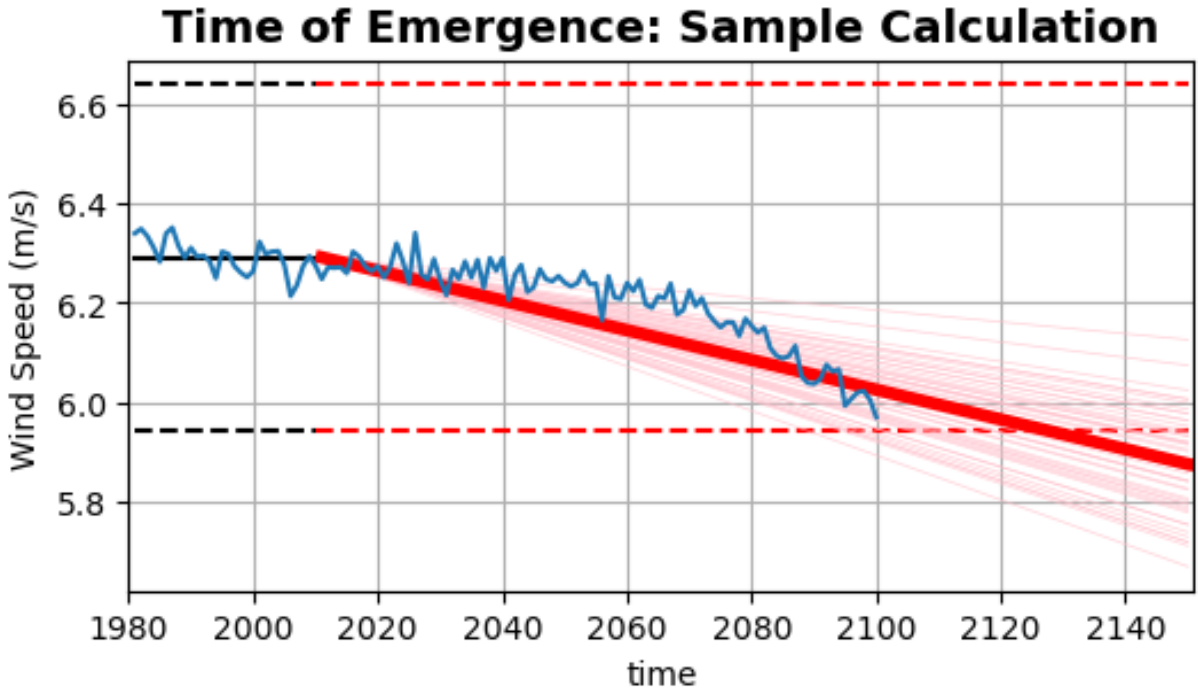

*Fig. S5 Example of the Time of Emergence (ToE) calculation. The blue line shows the CESM2 ensemble mean of the annual mean wind averaged in the climate zone of North Europe. The examined simulation years range from 1981 to 2100, and the period 1981–2010 is considered as the modern reference climate. The black coloring of straight lines denote 1981–2010, and the red coloring denote the years after 2010. The horizonal solid line shows the 30-year climate values in the reference period. The horizonal dashed lines show the two-standard-deviation range estimated using the mean of standard deviations of individual ensemble members during 1981–2010. Linear regression is conducted for the ensemble mean of 2011–2100 to estimate the long-term trend. The regression (thick red line) helps estimate the time when the ensemble mean may intersect with the two-standard-deviation threshold. In this case, the intersection occurs around Year 2125, so the year is considered as the ToE for the climate change signals to emerge from natural variability. The same regression is conducted for individual ensemble members (thin red lines) to estimate the uncertainty of the ToE.*

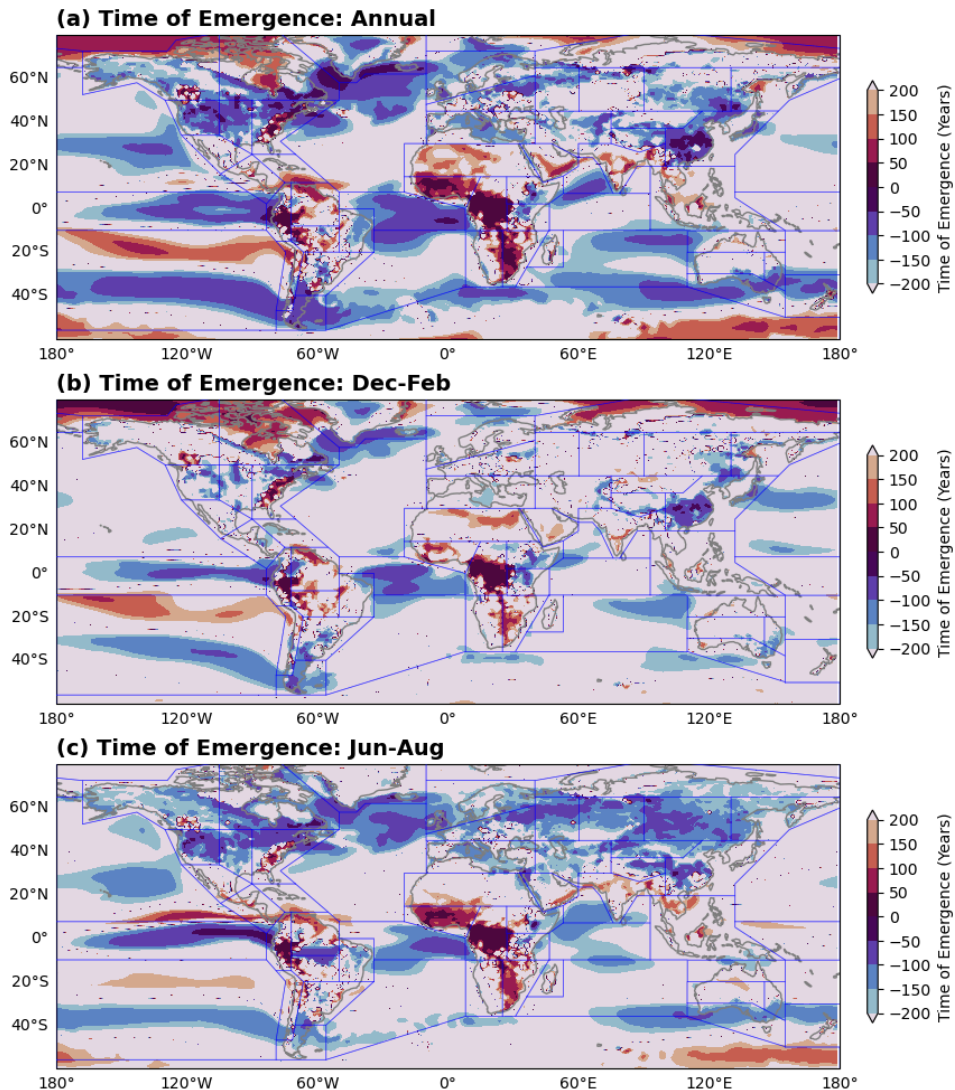

*Fig. S6 Time of emergence (TOE) for detectable human-contributed changes in near-surface wind speed to emerge. (a) TOE for changes in the annual mean. (b) TOE for changes in the December–February mean. (c) TOE for changes in the June–August mean. The TOE is indicated with the number of years after 2010. For example, a +50-year value indicates the signal of an increase in wind speed will appear by 2060, and -40-year value indicates the signal of a decrease in wind speed will appear by 2050. The blue polygons show the same climate zones as in Figure 1. The input data is the CESM2 large ensemble data.*

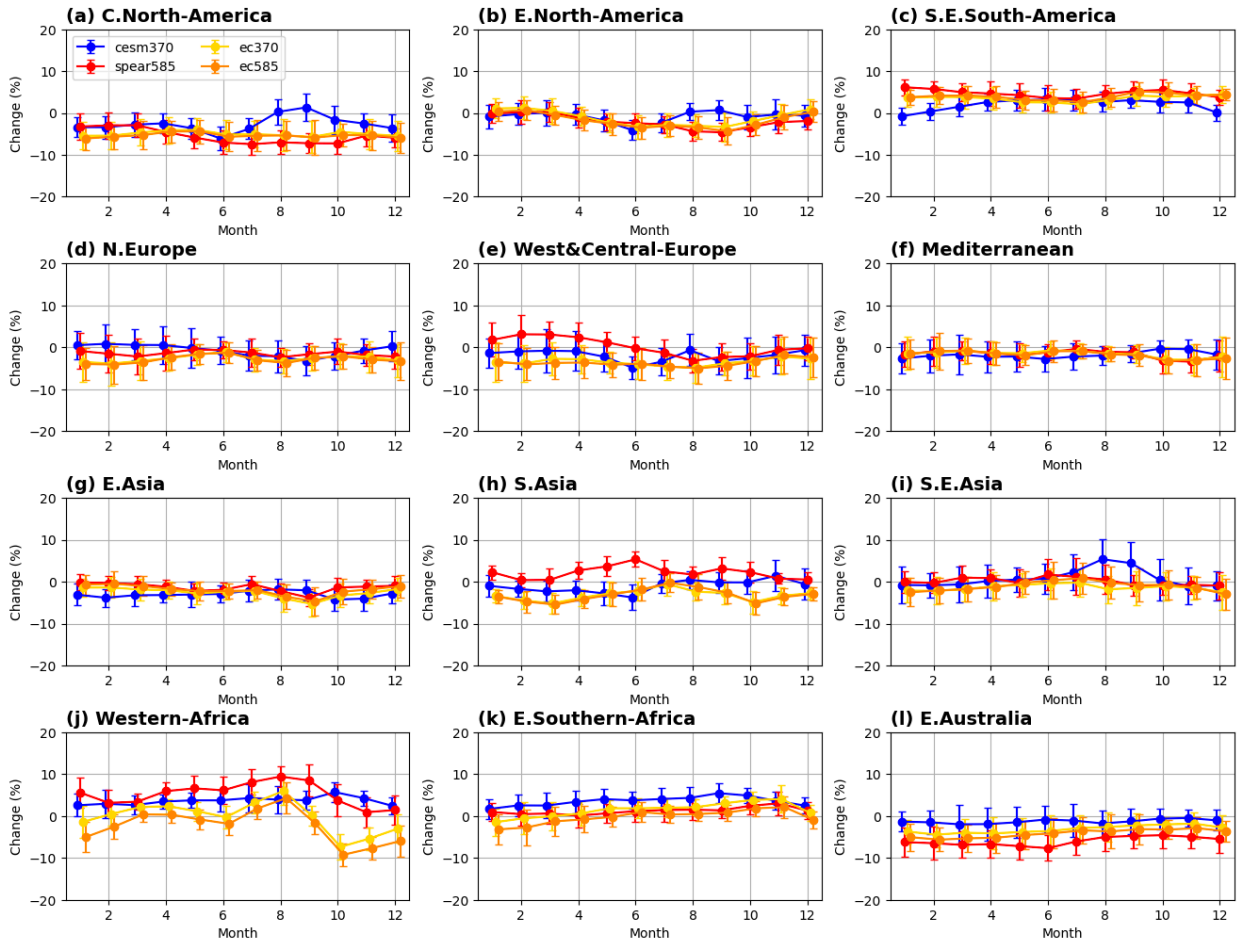

*Fig. S7 Relative Changes in near-surface wind speed projected by large ensemble simulations (2021–2050). The analyzed large ensemble simulations include the CESM2 simulations forced by the SSP3-7.0, the SPEAR simulations forced by the SSP5-8.5, and the EC-Earth simulations forced by the SSP3-7.0 and the SSP5-8.5. Regional averages are calculated based on the individual climate zones following Figure 1. In each set of the large ensemble simulations, the error bars indicate the range of ±2 standard deviations, and the reference climate is the historical simulations (1981–2010).*

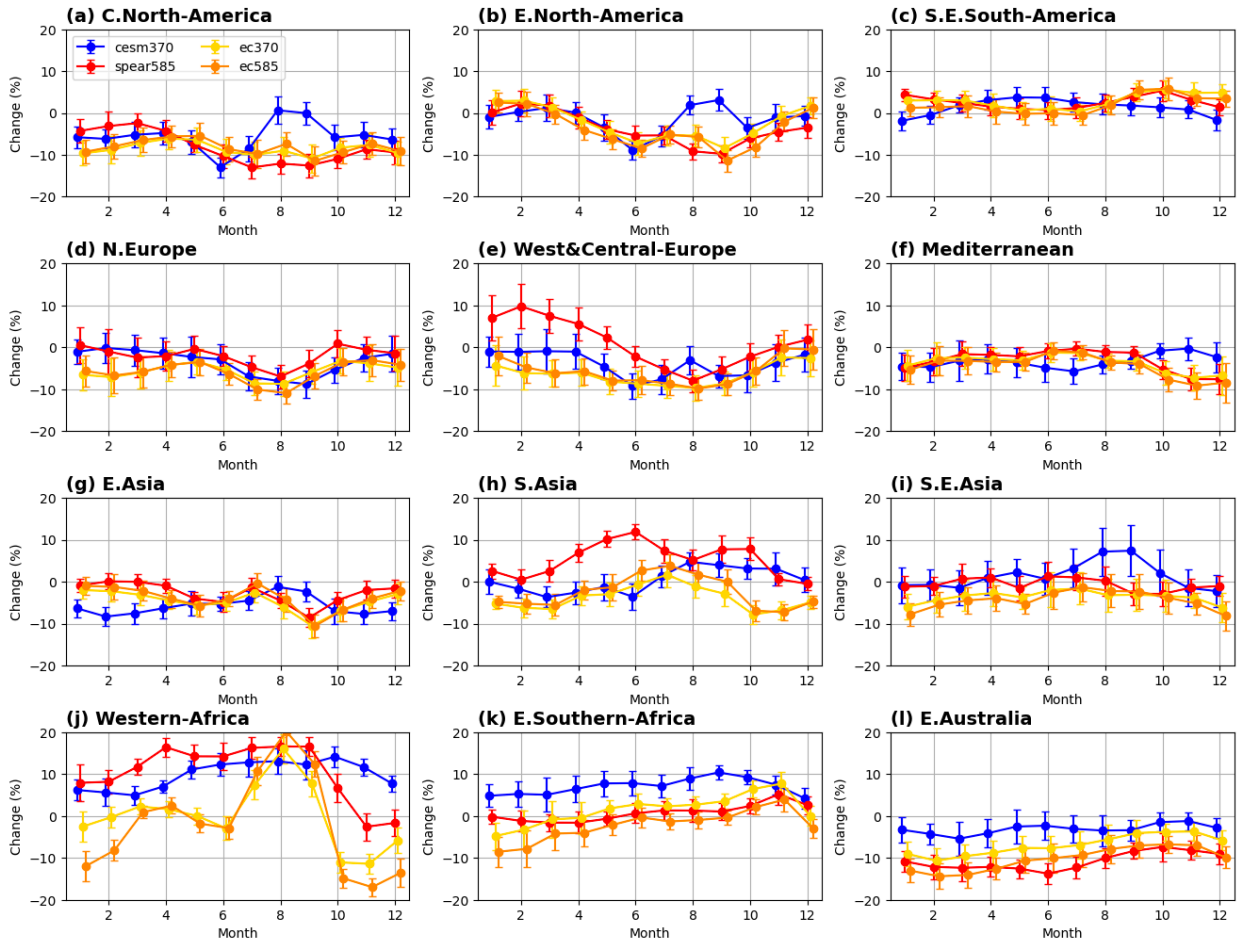

*Fig. S8 Relative Changes in near-surface wind speed projected by large ensemble simulations (2071–2100). The analyzed large ensemble simulations include the CESM2 simulations forced by the SSP3-7.0, the SPEAR simulations forced by the SSP5-8.5, and the EC-Earth simulations forced by the SSP3-7.0 and the SSP5-8.5. Regional averages are calculated based on the individual climate zones following Figure 1. In each set of the large ensemble simulations, the error bars indicate the range of ±2 standard deviations, and the reference climate is the historical simulations (1981–2010).*

NCAR-CESM CDD [18 degC]: SSP-370

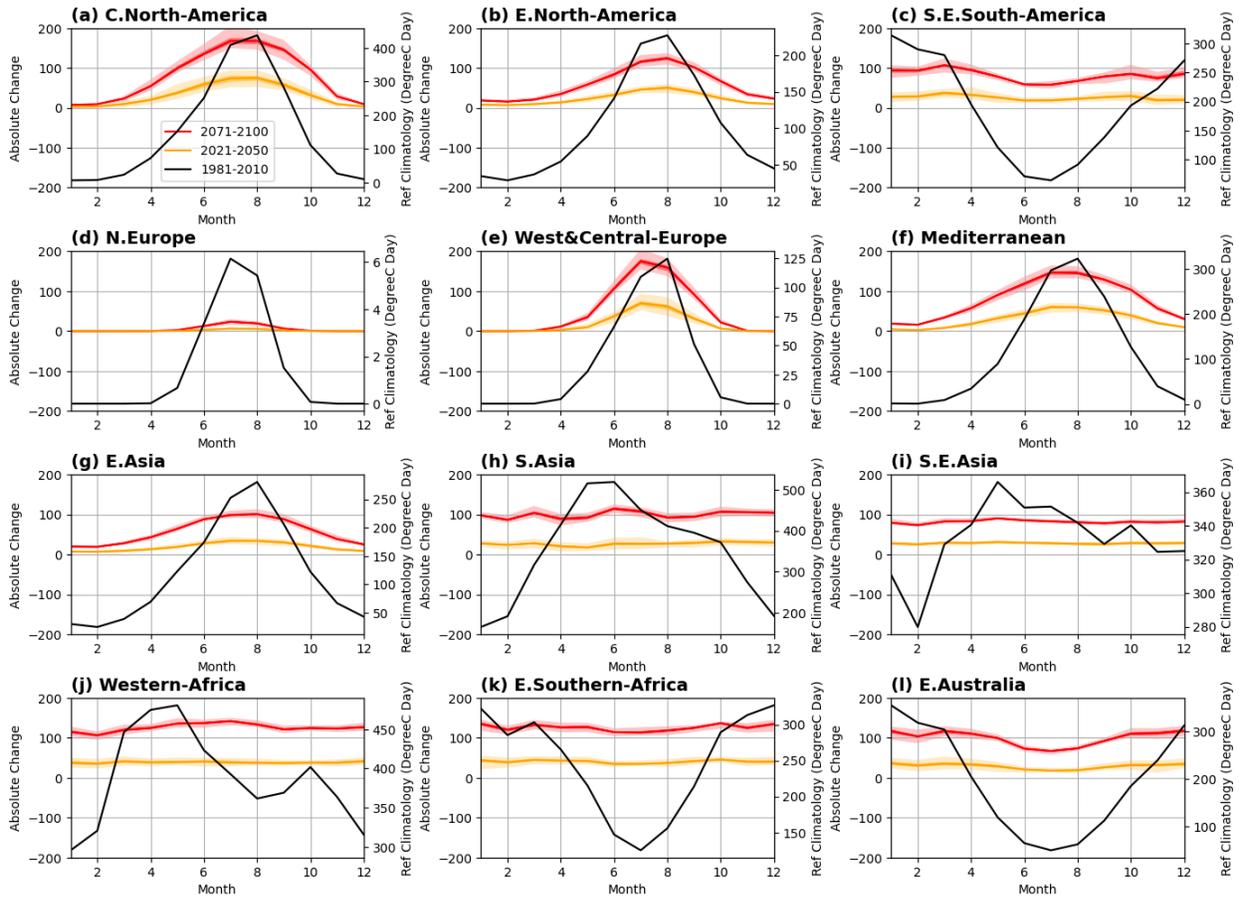

*Fig. S9 Absolute changes in cooling degree day ($T_{ref}$=18ºC) projected by the CESM2 large ensemble simulations. Regional averages are calculated based on the individual climate zones following Figure 1. Black line shows the annual mean cooling degree days during 1981–2010. Red and yellow lines indicate absolute changes in 2071–2100 and 2021–2050. The light shading shows the range of 0.5th and 99.5th percentiles, while the dark shading shows the range of 25th and 75th percentiles. The cooling degree day (ºC day) is calculated using the CESM2 large ensemble data.*

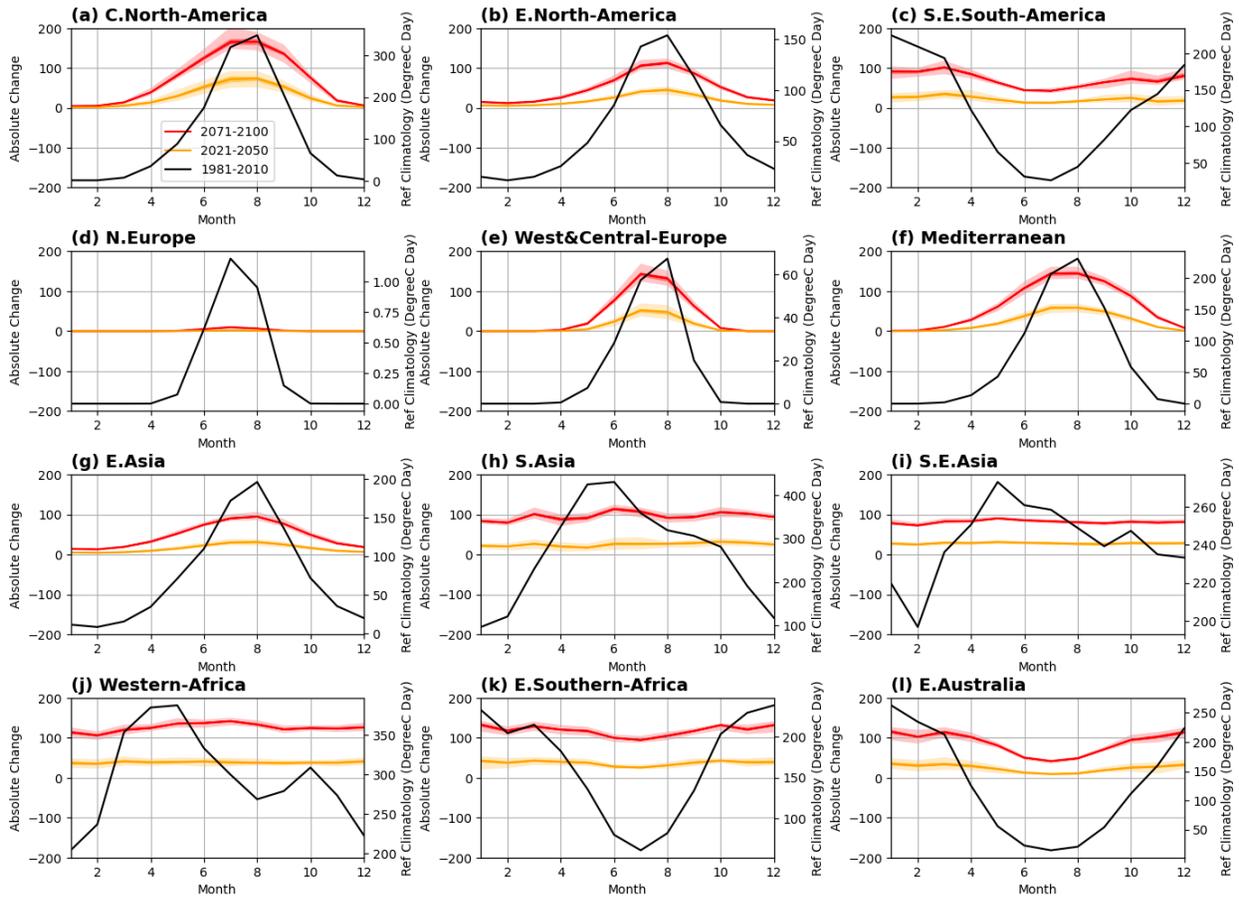

*Fig. S10 Absolute changes in cooling degree day ($T_{ref}$=21ºC) projected by the CESM2 large ensemble simulations. Regional averages are calculated based on the individual climate zones following Figure 1. Black line shows the annual mean cooling degree days during 1981–2010. Red and yellow lines indicate absolute changes in 2071–2100 and 2021–2050. The light shading shows the range of 0.5th and 99.5th percentiles, while the dark shading shows the range of 25th and 75th percentiles. The cooling degree day (ºC day) is calculated using the CESM2 large ensemble data.*

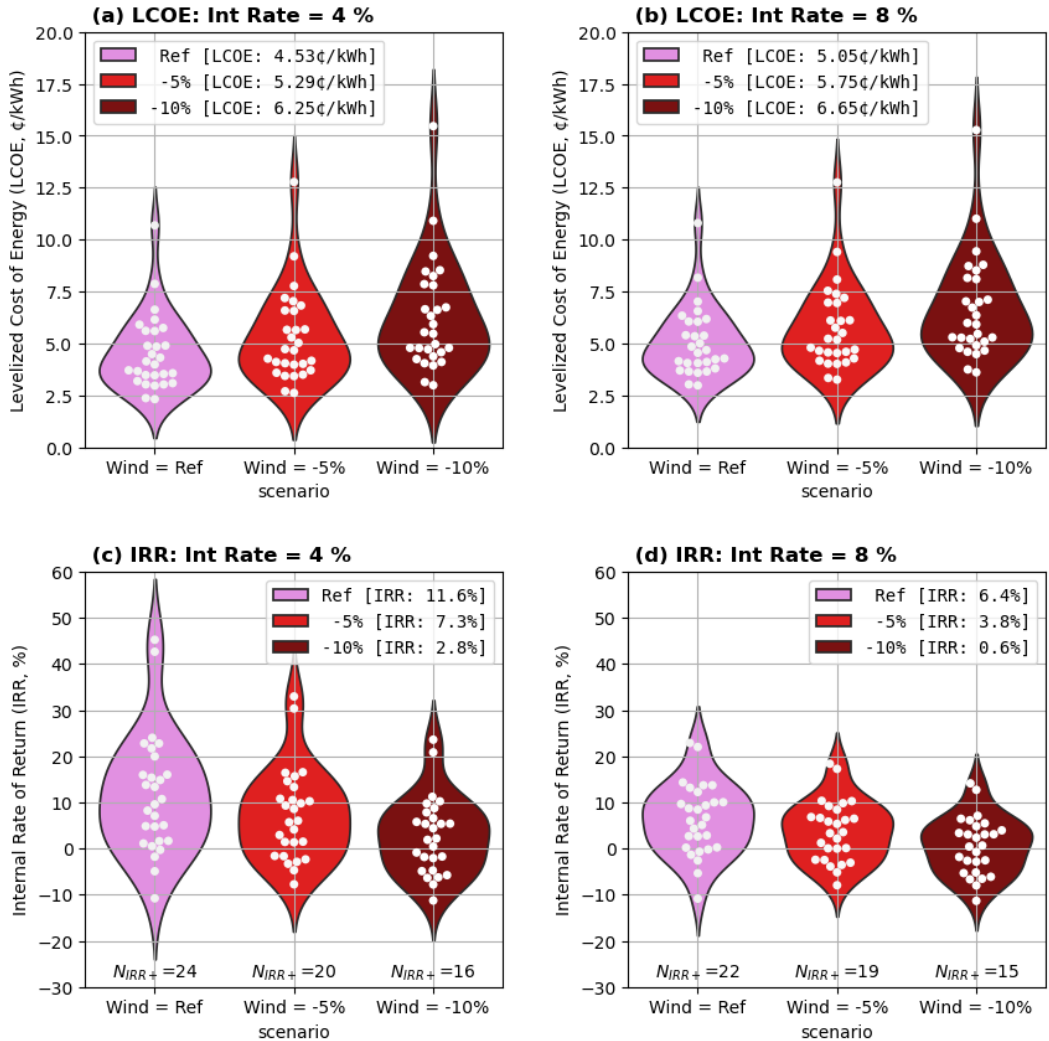

*Fig. S11 Impacts of wind changes and interest rates on the levelized cost of energy (LCOE) and the internal return rate (IRR) of wind energy projects. (a) The LCOE when wind speed remains unchanged (ref), reduces by 5%, and reduces by 10% when the interest rate holds at 4%. The LCOE at individual sites are denoted with white dots. The violin plots show the kernel densities of the LCOE. The LCOE average of all the sites is denoted in the legends. (b) Same as (a), but when the interest rate is at 8%. (c–d) Similar to (a–b), but for the IRR. Additionally, the numbers of sites with positive IRR are denoted at the bottom of violin plots. The evaluated sites are 28 representative onshore wind farm sites in the US. More details about the LCOE, IRR calculations, and wind farm sites are available in "Modeling of Wind Farm Cost-Competitiveness" section.*

*Table S1 Overview of the wind farm sites used in the cost competitiveness modeling. The location is represented with the subregions of individual US states since the NREL withheld the exact location to protect the interest of data providers.*

| Location | Topography | Location | Topography |
|---|---|---|---|
| Arkansas Northwestern | Flat Lands | Minnesota Southwestern | Flat Lands |
| Arizona Eastern | Rolling Hills | Montana Northwestern | Flat Lands |
| California Northern | Rolling Hills | North Dakota Northern | Flat Lands |
| California Southern | Mountainous | New Mexico Eastern | Flat Lands |
| California Southwestern | Mountainous | New York Northern | Flat Lands |
| Colorado Northeastern | Flat Lands | Ohio Northern | Lake |
| Colorado Southeastern | Flat Lands | Oregon Northern | Flat Lands |
| Florida Southern | Flat Lands | Texas Northwestern | Flat Lands |
| Idaho Southeastern | Mountainous | Texas Southeastern | Flat Lands |
| Indiana Northwestern | Flat Lands | Texas Southwestern | Flat Lands |
| Kansas Central | Flat Lands | Utah Southwestern | Flat Lands |
| Maine Northern | Flat Lands | Washington Central | Rolling Hills |
| Michigan Eastern | Flat Lands | West Virginia Northeastern | Rolling Hills |
| Michigan Western | Lake | Wyoming Southern | Flat Lands |